\documentclass[10pt]{iopart}
\pdfoutput=1 
\usepackage{graphicx}
\usepackage{amssymb}
\usepackage[colorlinks,linkcolor=black,urlcolor=black,citecolor=phthaloblue]{hyperref}
\usepackage{color}
\definecolor{phthaloblue}{rgb}{0.0, 0.06, 0.54}

\eqnobysec

\newcommand{\beq}{\begin{equation}}
\newcommand{\eeq}{\end{equation}}
\newcommand{\be}{\begin{equation*}}
\newcommand{\ee}{\end{equation*}}
\newcommand{\beqa}{\begin{eqnarray}}
\newcommand{\eeqa}{\end{eqnarray}}
\newcommand{\stackunder}[2]{\mathrel{\mathop{#1}\limits_{#2}}}
\newcommand{\abs}[1]{\vert#1\vert}
\newcommand{\bigabs}[1]{\left\vert#1\right\vert}
\newcommand{\ct}{{\rm cont}}
\newcommand{\cum}[1]{\mean{\!\mean{#1}\!}}
\newcommand{\dd}{{\rm d}}
\newcommand{\eff}{{\rm eff}}
\newcommand{\eps}{{\varepsilon}}
\newcommand{\erf}{\mathop{\rm erf}}
\newcommand{\erfc}{\mathop{\rm erfc}}
\newcommand{\ii}{{\rm i}}
\newcommand{\lam}{\lambda}
\newcommand{\lap}[1]{\mathrel{\mathop{\cal L}\limits_{#1}^{}}}
\renewcommand{\max}{\mathop{\rm max}}
\newcommand{\mean}[1]{\langle#1\rangle}
\newcommand{\prob}{\mathbb{P}}
\newcommand{\sg}{{\rm sing}}
\newcommand{\sone}[2]{\left[#1\atop#2\right]}
\newcommand{\stwo}[2]{\left\{#1\atop#2\right\}}
\newcommand{\sm}[1]{{\w M^{(#1)}}}
\newcommand{\sr}[1]{{\w R^{(#1)}}}
\newcommand{\var}{\mathop{\rm Var}\nolimits}
\newcommand{\w}[1]{{\tilde{#1}}}
\newcommand{\F}[1]{{\widehat{#1}}}
\newcommand{\Kst}{K_{\rm st}}
\newcommand{\Li}{{\rm Li}}
\renewcommand{\Re}{\mathop{{\rm Re}}}
\renewcommand{\Im}{\mathop{{\rm Im}}}
\newcommand{\rz}{\breve{z}}

\begin{document}

\title{Maximum and records of random walks with stochastic resetting}

\author{Claude Godr\`eche and Jean-Marc Luck}

\address{Universit\'e Paris-Saclay, CNRS, CEA, Institut de Physique Th\'eorique,
91191~Gif-sur-Yvette, France}

\begin{abstract}
We revisit the statistics of extremes and records of symmetric random walks with stochastic resetting,
extending earlier studies in several directions.
We put forward a diffusive scaling regime
(symmetric step length distribution with finite variance,
weak resetting probability)
where the maximum of the walk and the number of its records up to discrete time $n$
become asymptotically proportional to each other for single typical trajectories.
Their distributions obey scaling laws ruled by a common two-parameter scaling function,
interpolating between a half-Gaussian and a Gumbel law.
The exact solution of the problem for the symmetric exponential step length distribution
and for the simple Polya lattice walk,
as well as a heuristic analysis of other distributions,
allow a quantitative study of several facets
of the statistics of extremes and records beyond the diffusive scaling regime.
\end{abstract}

\address{\today}

\eads{\mailto{claude.godreche@ipht.fr}, \mailto{jean-marc.luck@ipht.fr}}

\maketitle

\section{Introduction}
\label{intro}

Recent years have seen a flowering of studies on stochastic processes with resetting.
In particular, in view of the paradigmatic role played by random walks or
Brownian motion in the field of stochastic processes and in statistical
physics, special attention has been paid to the study of these simple processes
in the presence of stochastic resetting.
A comprehensive bibliography may be found in~\cite{emsrev}.
Virtually all the many facets of the study of random walks and Brownian motion
without resetting have an interesting counterpart in the presence of resetting.
One may think for instance of first-passage problems, or of the statistics of extremes and records,
to name but a few.

Very recently, the statistics of records for random walks
with stochastic resetting has been addressed in~\cite{m2s2}.
The central result of that work is an expression for the mean number of records,
which is universal, i.e., independent of the step length distribution,
provided the latter is symmetric and continuous.

The purpose of the present work is to revisit the problem
and to extend the results of~\cite{m2s2} and those coming from earlier studies~\cite{em,mmss}
in several directions.
Consider for definiteness a random walk starting from the origin,
defined by the recursion
\beq
x_{n+1}=\left\{
\matrix{
0\hfill &\hbox{with prob.~$r$},\hfill\cr
x_n+\eta_{n+1}\quad &\hbox{with prob.~$1-r$}.
}
\right.
\label{def}
\eeq
At each time step, the walker is reset to the origin with probability $r$.
The step lengths~$\eta_n$ have a symmetric distribution
with density $\rho(\eta)$ and finite variance
\beq\label{variance}
\mean{\eta_1^2}=\int_{-\infty}^\infty\eta^2\rho(\eta)\dd\eta=\sigma^2=2D,
\eeq
where $D$ is the diffusion coefficient.
In most of this paper, the step length distribution is assumed to be continuous.
We shall address the case of discrete distributions
in section~\ref{simple}, when considering the simple Polya walk.

Throughout this work,
we consider the following two quantities in parallel.
The first one is the maximal height attained by the walk after $n$ steps,
\beq
M_n=\max(0,x_1,\dots,x_n).
\eeq
The second quantity of interest is the number $R_n$ of records up to time $n$.
We recall that a record occurs at step $m$
if $x_m$ is larger than all the previous positions of the walk $(0,x_1,\dots,x_{m-1})$\footnote{We refer the reader to~\cite{revue} for a recent review on records for random walks and L\'evy flights.}.
We shall present two different kinds of results
concerning the statistics of $M_n$ and~$R_n$ for symmetric random walks with resetting.

First, we put forward a diffusive scaling regime
(symmetric step length distribu\-tions with finite variance,
long walks and weak resetting probability),
for which the asymptotic equivalence
\beq
M_n\approx E\sqrt{D}\,R_n
\label{mrintro}
\eeq
holds for single typical trajectories,
implying in particular that the distributions of $M_n$ and~$R_n$
are simply related to each other all over this regime.
The enhancement factor $E\ge1$, given by~(\ref{edef}),
is unity for continuous distributions,
and larger than unity for discrete distributions,
and more generally distributions having a discrete component.
The scaling forms of the distributions of $M_n$ and $R_n$ read
\beq\label{eq:equivscal}
f_n(M)\approx\frac{1}{\sqrt{Dn}}\,\Phi(X,u),\qquad
p_n(R)\approx\frac{E}{\sqrt{n}}\,\Phi(X,u),
\eeq
where the reduced distribution $\Phi(X,u)$, whose analytical expression is given in~(\ref{phidef}),
is a scaling function of the two variables
\beq
X=\frac{M}{\sqrt{Dn}}\quad\hbox{or}\quad X=\frac{ER}{\sqrt{n}},\qquad u=nr,
\eeq
interpolating between a half-Gaussian law for $u\ll1$ and a Gumbel law for $u\gg1$.
Figure~\ref{un} confirms the predictions~(\ref{mrintro}) and~(\ref{eq:equivscal})
by depicting the excellent asymptotic agreement
between the exact theoretical expression~(\ref{phidef})
of the reduced distribution $\Phi(X,u)$ and the rescaled distribution of $R_n$ obtained by simulations,
on the example of a walk with uniformly distributed step lengths.

In a second part of this work, we investigate several facets of the problem
beyond the above diffusive scaling regime, as detailed in the outline of the paper which follows.

Section~\ref{gal} is devoted to some general formalism,
where we derive a fundamental integral equation for the distribution of $M_n$,
we recall the Wiener-Hopf approach,
and relate the distribution of $M_n$ with resetting
to the same distribution without resetting.
The core of the analysis of the diffusive scaling regime is done in section~\ref{sca}.
We rely on the renewal structure of the record process to establish the domain of validity
of the equivalence~(\ref{mrintro}), first without resetting, then with resetting
(section~\ref{equiv}).
The key scaling formula~(\ref{key}) for the distributions of $M_n$ and~$R_n$
is derived in section~\ref{scakey}
and exploited in detail in sections~\ref{scamoms} (moments)
and~\ref{scafull} (full distributions), where the reduced distribution $\Phi(X,u)$ is investigated at depth.

Section~\ref{beyond} contains several results
illustrating to what extent the statistics of $M_n$ and of $R_n$ may differ
for generic symmetric step length distributions.
In section~\ref{wo} we analyze the leading corrections to the scaling behavior
of the mean values $\mean{M_n}$ and $\mean{R_n}$ without resetting,
whereas section~\ref{gumregime} is devoted to the regime of late times at fixed non-zero
resetting probability.
For superexponential step length distributions,
both $M_n$ and $R_n$ grow logarithmically in time,
and the corresponding amplitudes are determined.
This is illustrated in figure~\ref{aruni} where the
mean values $\mean{M_n}$ and $\mean{R_n}$ are plotted against $\ln n$,
for a uniform step length distribution and a finite resetting
probability $r=1/2$, demonstrating clearly that the asymptotic equivalence~(\ref{mrintro})
does not hold beyond the diffusive scaling regime.
For subexponential step length distributions,
there is a qualitative difference between the growth laws of $M_n$ and $R_n$, as
corroborated by numerical simulations shown in figure~\ref{artheta}.

Finally, two examples of step length distributions for which the fundamental integral equation
can be solved exactly by elementary means are considered in section~\ref{special}.
For each of these examples, we first show how universal results
are recovered in the diffusive scaling regime,
and then investigate some specific features beyond that regime.
For the symmetric exponential distribution (section~\ref{se}),
we demonstrate the existence of an infinity of linear identities between the moments of $M_n$ and $R_n$
in the absence of resetting.
In the case of the binary distribution yielding the simple Polya walk (section~\ref{simple}),
we have $M_n=R_n$ identically.
We revisit the analysis of $\mean{M_n}$ in the absence of resetting,
and make an excursion into the realm of generic discrete distributions,
and more generally distributions having a discrete component.
We then characterize the logarithmic growth of $\mean{M_n}$
in the presence of resetting,
calculate its amplitude,
and show that it is modulated by periodic oscillations,
that are also determined.
We come back to the most salient results of the present work
in a brief discussion (section~\ref{disc}).
\ref{quick} gives an alternative derivation of some of the results of~section~\ref{se},
based on the renewal structure of the record process,
whereas three other appendices contain more technical material.

\section{General formalism}
\label{gal}

Throughout this work
we consider the random walk in discrete time with stochastic resetting
defined by the recursion~(\ref{def}).
Except in section~\ref{simple},
the step length distribution is assumed to be continuous and symmetric.

Let us first introduce some notations.
We denote the distribution function of the continuous random variable~$M_n\ge0$,
the maximal height of the walk after $n$ steps, by
\beq
F_n(M)=\prob(M_n\le M),
\label{Fdef}
\eeq
and by $f_n(M)=F_n'(M)$ the corresponding density, such that
\beq
f_n(M)\dd M=\prob(M<M_n<M+\dd M).
\label{fdef}
\eeq
The number of records $R_n$ up to time $n$
is a discrete random variable, whose distribution will be denoted as
\beq
p_n(R)=\prob(R_n=R),
\label{pdef}
\eeq
where $R\ge0$ is an integer.

The main purpose of this section is to show how
the distribution of the maximum~$M_n$ can be determined in full generality.
Consider a random walk of $n$ steps starting from an arbitrary initial position $x_0=x$.
For any fixed height $M\ge0$, we denote by
\beqa
Q_n(x,M)&=&\prob(M_n\le M|x_0=x)
\nonumber\\
&=&\prob(x_0\le M,\dots,x_n\le M|x_0=x)
\label{qdef}
\eeqa
the probability that its maximum $M_n$ up to time $n$ is at most $M$.
We have in particular
\beq
F_n(M)=Q_n(0,M).
\label{fq}
\eeq
The probability $Q_n(x,M)$ is non-zero only for $x\le M$.
In this range, it obeys the backward integral equation
\beq
Q_{n+1}(x,M)=rQ_n(0,M)+(1-r)\int_{-\infty}^{M-x}Q_n(x+\eta,M)\rho(\eta)\dd\eta,
\eeq
with initial condition $Q_0(x,M)=1$.
This equation is obtained by conditioning on the first step of the walk,
which may be a resetting event or not.
The generating series
\beq
\w Q(z,x,M)=\sum_{n\ge0}Q_n(x,M)z^n
\label{qser}
\eeq
therefore obeys
\beq
\w Q(z,x,M)=1+rz\w Q(z,0,M)
+\rz\int_{-\infty}^{M-x}\w Q(z,x+\eta,M)\rho(\eta)\dd\eta.
\label{qeq}
\eeq
Here and throughout this paper, we use the shorter notation
\beq
\rz=(1-r)z.
\label{rzdef}
\eeq
We shall now see that the integral equation~(\ref{qeq}) can be solved by Wiener-Hopf techniques
for an arbitrary symmetric continuous step length distribution.

\subsubsection*{In the absence of resetting.}

For $r=0$,~(\ref{qeq}) becomes invariant under a simultaneous translation of $x$ and $M$.
Its solution therefore reads
\beq
\w Q(z,x,M)=\w q(z,M-x),
\label{qg}
\eeq
where $\w q(z,y)$ is the generating series of the probability $q_n(y)$
that a walker starting from position $y\ge0$ does not cross the origin up to time $n$,
\beq
q_n(y)=\prob\left(x_1\ge0,\dots,x_n\ge0|x_0=y\right).
\eeq
This survival probability coincides with the distribution function of the maximum,
\beq
q_n(y)=F_n(y)=\prob(M_n\le y).
\label{eq:coincide}
\eeq
The generating series $\w q(z,y)$ obeys the inhomogeneous Milne integral equation with a constant source term,
\beq
\w q(z,y)=1+z\int_0^\infty\w q(z,y')\rho(y-y')\dd y'\qquad(y\ge0),
\label{geq}
\eeq
which can be solved by means of the Wiener-Hopf factorization~\cite{hopf,chandra}
(see~\cite{LA} for a historical account of this method).
The solution reads in Laplace space
\beq
\fl
\int_0^\infty\w q(z,y)\,\e^{-py}\,\dd y
=\frac{1}{p\sqrt{1-z}}\,
\exp\left(-\frac{p}{\pi}\int_0^\infty\frac{\ln(1-z\F\rho(k))}{p^2+k^2}\,\dd k\right),
\label{lres}
\eeq
where
\beq
\F\rho(k)=\int_{-\infty}^\infty\e^{-\ii k\eta}\rho(\eta)\dd\eta
\label{foudef}
\eeq
is the Fourier transform of the step length distribution.
The result~(\ref{lres}) is a variant of the Pollaczek-Spitzer formula
(see~\cite{spitzer1,spitzer2,ivanov}).
We have in particular
\beq
\w q(z,0)=\frac{1}{\sqrt{1-z}},\qquad
\lim_{y\to\infty}\w q(z,y)=\frac{1}{1-z}.
\label{qlims}
\eeq
Using the shorter notations $\w q(z)$ for $\w q(z,0)$ and $q_n$ for $q_n(0)$,
we recover the well-known result of Sparre Andersen theory~\cite{sparre53,sparre54}
(see~\cite[chap.~XII]{feller2} for a simple presentation) for the probability $q_n$ that a
walker starting from the origin does not cross the origin up to time $n$,
\beq
\w q(z)=\sum_{n\ge0}q_nz^n=\frac{1}{\sqrt{1-z}},
\label{qzres}
\eeq
i.e.,
\beq
q_n=b_n,
\label{qnres}
\eeq
where $b_n$ is the binomial probability
\beq
b_n=\frac{(2n)!}{(2^nn!)^2}=\frac{{2n\choose n}}{2^{2n}}.
\label{bdef}
\eeq
The expressions~(\ref{qzres}) and~(\ref{qnres}) are universal, i.e.,
independent of the step length distribution,
as long as it is symmetric and continuous, with finite or infinite variance.

\subsubsection*{In the presence of resetting.}

The mere linearity of~(\ref{qeq})
implies that its solution $\w Q(z,x,M)$ in the presence of resetting
can be expressed in terms of its solution $\w q(z,y)$ in the absence of resetting,
i.e.,
\beq
\w Q(z,x,M)=(1+rz\w Q(z,0,M))\,\w q(\rz,M-x).
\eeq
Determining self-consistently the quantity $\w Q(z,0,M)$ from this equation, we obtain
\beq
\w Q(z,x,M)=\frac{\w q(\rz,M-x)}{1-rz\w q(\rz,M)}.
\label{qfull}
\eeq
The connection with the distribution of the maximum $M_n$ is given
by~(\ref{fq}), i.e.,
\beq
\w F(z,M)=\w Q(z,0,M)=\frac{\w q(\rz,M)}{1-rz\w q(\rz,M)},
\label{fqser}
\eeq
which can be more easily remembered in the form
\beq
\w F(z,M)|_{r}=\frac{\w F(\rz,M)|_{0}}{1-rz\w F(\rz,M)|_{0}},
\label{fqser+}
\eeq
with the notation
\beq
\w F(z,M)|_{0}=\w q(z,M),
\label{fqzero}
\eeq
in agreement with~(\ref{eq:coincide}).

Hereafter we shall only need the solution of the integral equation~(\ref{qeq})
in several specific situations where it can be solved by elementary means,
including the diffusive scaling regime (see~section~\ref{scakey}),
the symmetric exponential step length distribution (see~section~\ref{se})
and the symmetric binary distribution,
yielding the simple random walk, or Polya walk, on the one-dimensional lattice
(see~section~\ref{simple}).

\section{Diffusive scaling regime}
\label{sca}

The purpose of this section is to describe in detail the universal scaling laws
which govern the statistics of the maximum $M_n$ and of the number $R_n$ of records
throughout the diffusive scaling regime
of long walks ($n\gg1$) in the presence of a weak resetting ($r\ll1$),
for an arbitrary continuous symmetric step length distribution with finite variance.
The case of discrete distributions will be discussed in section~\ref{simple},
devoted to the simple Polya walk.

\subsection{A remarkable asymptotic equivalence}
\label{equiv}

We begin by putting forward the remarkable asymptotic equivalence
\beq
M_n\approx\sqrt{D}\,R_n.
\label{mrequiv}
\eeq
As will be shown below, this equivalence holds for single typical trajectories,
for all continuous symmetric step length distributions with finite variance.
It implies in particular that the asymptotic distributions of $M_n$ and~$R_n$
are simply related to each other.

\subsubsection*{In the absence of resetting.}

\begin{figure}[!ht]
\begin{center}
\includegraphics[angle=0,width=.48\linewidth,clip=true]{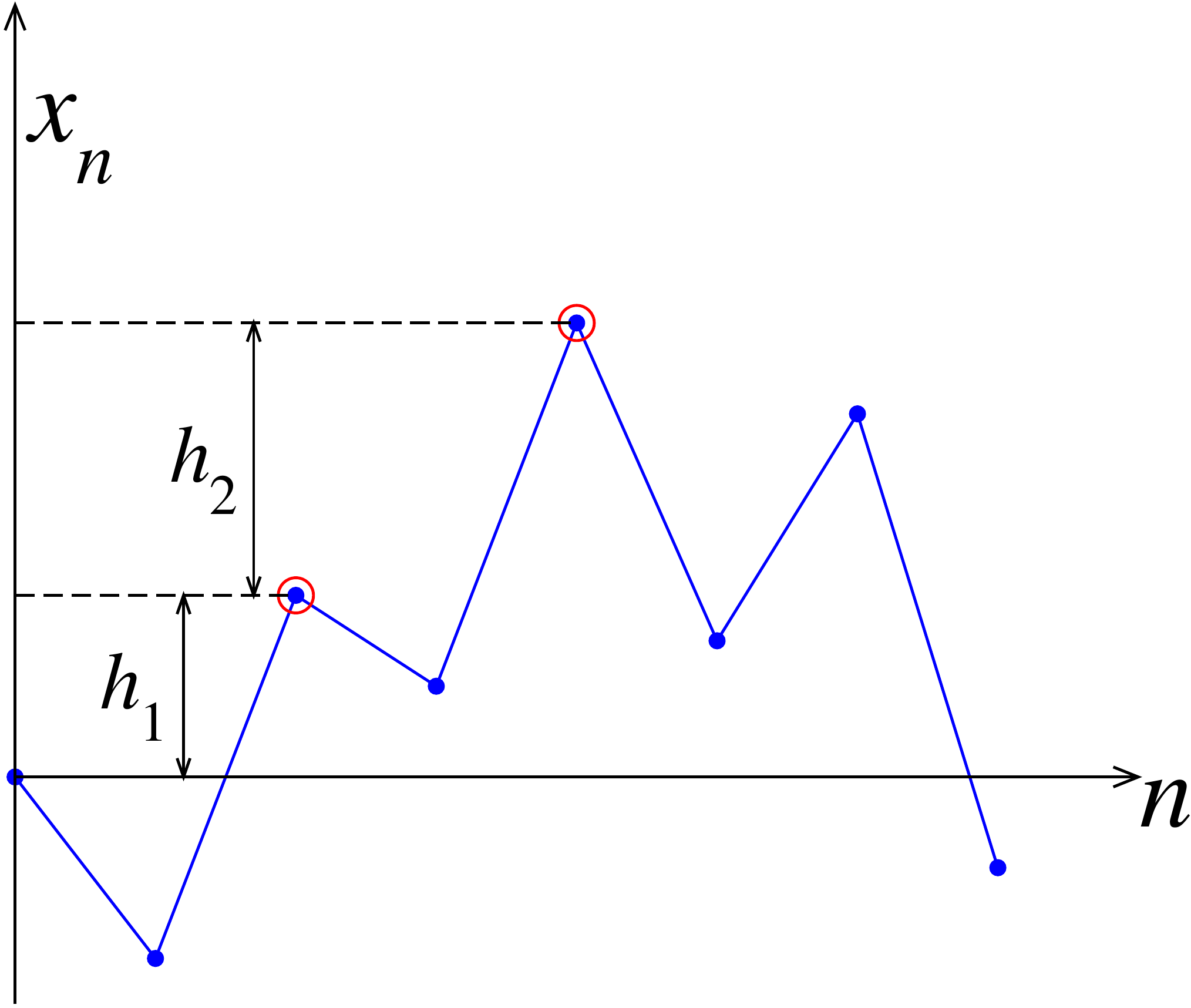}
\hskip 8pt
\includegraphics[angle=0,width=.48\linewidth,clip=true]{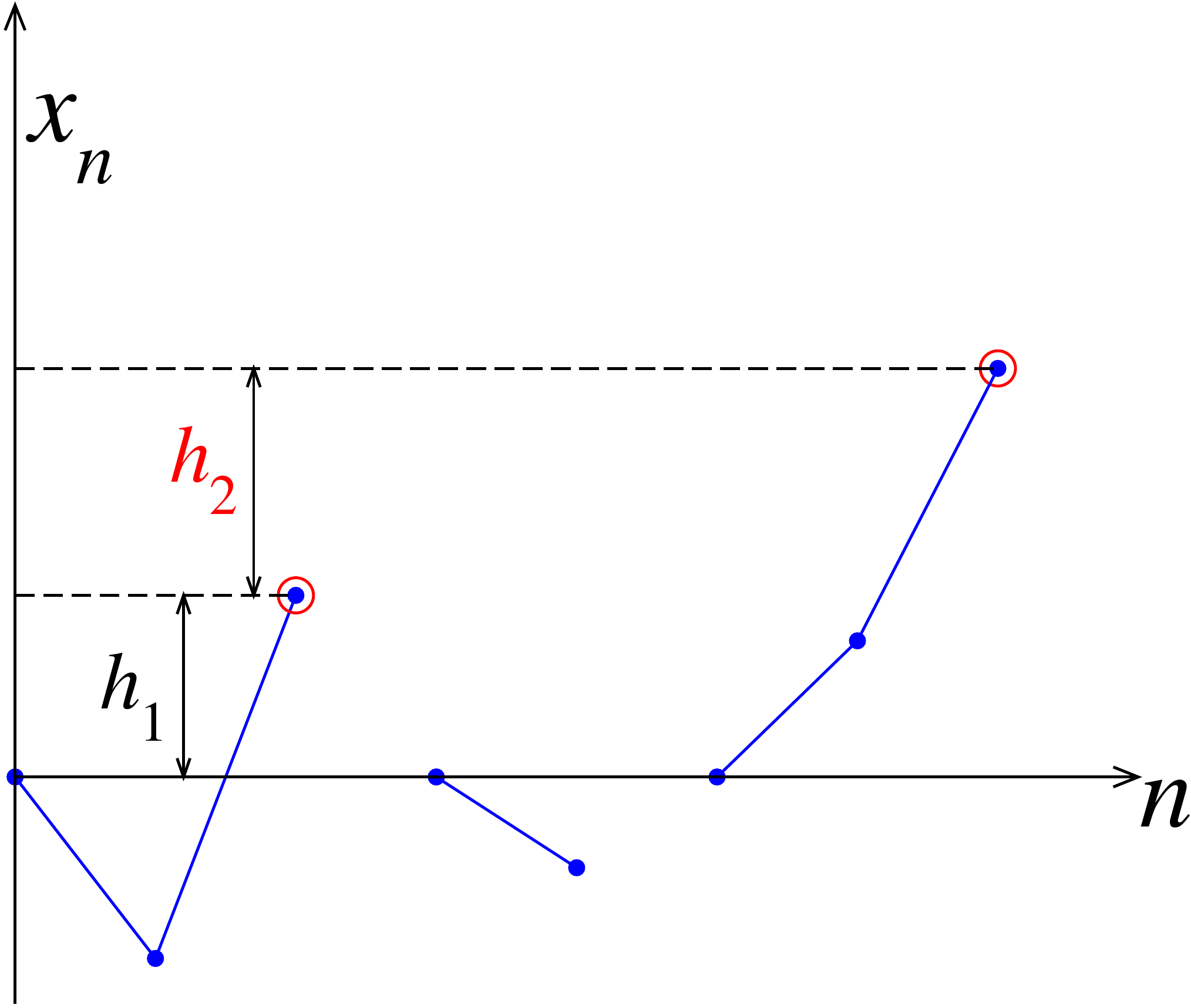}
\caption{\small
Schematic picture of a random walk,
showing records and increments between record positions.
Left: in the absence of resetting,
the increments $h_1$ and $h_2$ are iid with common distribution $f_{h_1}(h)$.
Right: in the presence of resetting,
the increment $h_1$ is distributed according to $f_{h_1}(h)$,
whereas the inter-resetting increment $h_2$ is not.}
\label{walk}
\end{center}
\end{figure}

In the absence of resetting, the equivalence~(\ref{mrequiv})
is a consequence of the renewal structure of the record process.
Let $n_k$ be the time of occurrence of the $k$th record and $x_{(k)}=x_{n_k}$
the corresponding position of the walker.
This renewal structure implies that the increments
\beq
h_k=x_{(k)}-x_{(k-1)}
\label{hdef}
\eeq
are independent and identically distributed (iid) random
variables~\cite{feller2,blackwell}.
Their common distribution, denoted by $f_{h_1}(h)$,
is also the distribution of the first positive position
of the random walk starting at the origin,
\beq
h_1\equiv x_{(1)},
\eeq
irrespective of the time $n_1$ at which this position is reached.
The left panel of figure~\ref{walk} gives an illustration of the process.

A remarkable result due to Spitzer~\cite{spitzer3,spitzerbook}
(see also~\cite[ch.~XVIII]{feller2}) states that the mean value of $h_1$ only
depends on the diffusion coefficient, as
\beq
\mean{h_1}=\sqrt{D},
\label{spitzer}
\eeq
for all continuous symmetric step length distributions with finite variance.

The maximum $M_n$ of the walk at time $n$ is the position of the current record, i.e.,
\beq\label{eq:Mn}
M_n=x_{(R_n)}=h_1+h_2+\cdots+h_{R_n}.
\eeq
In the regime of late times,
the number of records $R_n$ is typically large, scaling as~$\sqrt{n}$.
The law of large numbers therefore tells us that it is legitimate to replace
each increment $h_k$ by its mean value, obtaining thus
\beq
M_n\approx\mean{h_1}\,R_n.
\label{mrequiv2}
\eeq
For continuous step length distributions,
the expression~(\ref{spitzer}) of $\mean{h_1}$
implies the equivalence~(\ref{mrequiv})
for every single typical trajectory.

Note that, for L\'evy flights with $\rho(\eta)\sim |\eta|^{-(1+\theta)}$ ($0<\theta<2$) the distribution of the increments $f_{h_1}(h)\sim h^{-(1+\theta/2)}$~\cite{gmsprl}, which rules out the law of large numbers in this case (see also (\ref{eq:Lev}) and section \ref{gumregime} for a more complete discussion).

For symmetric discrete distributions,
and more generally symmetric distributions having a discrete component,
we mention in anticipation of the discussion given in section~\ref{simple}
that the Spitzer formula~(\ref{spitzer}) generalizes to
\beq
\mean{h_1}=E\sqrt{D},
\label{spitzeranti}
\eeq
where the enhancement factor $E\ge1$ depends on the step length distribution
according to~(\ref{edef}).
The asymptotic equivalence~(\ref{mrequiv}) therefore becomes
\beq
M_n\approx E\sqrt{D}\,R_n.
\label{mrequivanti}
\eeq
The enhancement factor $E$ is unity for all continuous symmetric distributions,
so that~(\ref{spitzer}) and~(\ref{mrequiv}) are recovered.

Coming back to continuous symmetric step length distributions,
the distribution $f_{h_1}(h)$ of the increments has been investigated
in~\cite{feller2,revue,gmsprl}, where
it is calculated explicitly for a few specific cases.
A general expression for this distribution
can be derived by means of Wiener-Hopf techniques~\cite{ustc}.
We have shown in particular that its second moment,
\beq
\mean{h_1^2}=2\sqrt{D}\,\ell,
\eeq
depends on details of the step length distribution,
because it involves the extrapolation length (or extrapolation distance) $\ell$.
The latter quantity has a long history,
going back to the works by Milne in radiative transfer theory~\cite{milne} (see also~\cite{mounaix17} for a short historical account).
It may be defined by considering the homogeneous Milne integral equation
\beq\label{milne}
H(y)=\int_0^\infty H(y')\rho(y-y')\dd y',
\eeq
which has a solution growing as
\beq
H(y)\approx y+\ell\qquad(y\to\infty).
\eeq
The following expressions of the extrapolation length,
\beq
\ell
=\frac{1}{\pi}\int_0^\infty\frac{1}{k^2}\,\ln\frac{Dk^2}{1-\F\rho(k)}\,\dd k
=\frac{1}{\pi}\int_0^\infty\frac{1}{k}\left(\frac{\F\rho\,'(k)}{1-\F\rho(k)}+\frac{2}{k}\right)\dd k,
\label{extra}
\eeq
are finite whenever $\mean{\abs{\eta_1^3}}$ is convergent~\cite{spitzer2}.
This condition is more stringent than the finiteness of the variance.
As in~(\ref{foudef}), $\F\rho(k)$ denotes the Fourier transform of $\rho(\eta)$.

As a further consequence of the renewal structure mentioned above,
the waiting times between two consecutive records, $n_1,n_2-n_1,\dots$ are iid
random variables with common distribution
\beqa
f_n=\prob(x_1\le0,\cdots,x_{n-1}\le0,x_n>0)=q_{n-1}-q_n,
\eeqa
which is the probability that the first entry of the walk on the positive side
occurs at the $n$th step.
We have thus $f_n=\prob(n_1=n)$ and, accordingly, $q_n=\prob(n_1>n)$.
The corresponding generating series reads
\beq
\w f(z)=\sum_{n\ge1}f_n z^n=1-(1-z)\w q(z)=1-\sqrt{1-z}
\label{fzres}
\eeq
(see~(\ref{qzres})).
We have therefore
\beq
f_n=\frac{b_n}{2n-1}\qquad(n\ge1),
\eeq
where $b_n$ is defined in~(\ref{bdef}).
This result is universal for walks with continuous symmetric step length distributions,
as is the expression~(\ref{qnres}) of $q_n$.

\subsubsection*{In the presence of a weak resetting.}

The asymptotic equivalence~(\ref{mrequiv}) still holds
in the presence of a weak stochastic resetting ($r\ll1$).
In this regime, the geometric distribution of the lapse of time $T$ between
successive resetting events,
\beq
\prob(T=j)=r(1-r)^{j-1}\qquad(j\ge1),
\label{geolaw}
\eeq
becomes very broad,
as testified by its mean value
\beq
\mean{T}=\frac{1}{r}.
\eeq

The derivation of the equivalence~(\ref{mrequiv}) given above has to be adapted
in two ways.
First, for a long walk with weak resetting,
the typical number $R_n$ of records is still large.
It is indeed at least as large as the number of records before the first
resetting event,
which scales as $\sqrt{\mean{T}}$, i.e., as $1/\sqrt{r}$.
Second, keeping the notation~(\ref{hdef}) for the increments between successive records
in the presence of resetting,
most increments~$h_k$ correspond to both records $x_{(k)}$ and $x_{(k-1)}$
belonging to the same stretch of random walk between two successive
resettings,
and are therefore distributed according to $f_{h_1}(h)$.
Inter-resetting increments $h_k$,
where the successive records $x_{(k)}$ and $x_{(k-1)}$ are separated by at
least one resetting,
are in general neither independent of the rest of the walk
nor distributed according to $f_{h_1}(h)$.
This is illustrated in the right panel of figure~\ref{walk}.
The number of such inter-resetting increments is however small, as it
is at most equal to the number of resettings up to time~$n$.
This number has a binomial distribution with mean value
\beq
u=nr.
\label{udef}
\eeq

We have therefore shown that the asymptotic equivalence~(\ref{mrequiv})
holds for $nr\ll1/\sqrt{r}$, i.e., $r\ll n^{-2/3}$.
In particular, it holds in the diffusive scaling regime
at any fixed value of the scaling variable $u$.
However, it does not hold in general for arbitrary values of the resetting probability $r$,
since a finite fraction of the increments $h_k$ are inter-resetting increments
whose distribution is not under control (see section~\ref{gumregime}).

Let us mention that the connection between the mean number of records and the average maximum of 
a random walk with resetting is also discussed in~\cite{m2s2}, as we now summarize.
In the absence of resetting, it is stated in~\cite{m2s2},
on the basis of arguments given in~\cite{wms}, that the asymptotic relation~(\ref{mrequiv})
holds for random walks with continuous jump distribution $\rho(\eta)$ and finite variance.
The relation~(\ref{mrequiv}) is then conjectured in~\cite{m2s2}
to be still valid in the presence of resetting.
On the one hand, it is stated that~(\ref{mrequiv}) should hold for a discrete-time
random walk with a continuous and symmetric jump distribution and finite variance
in the weak resetting regime. 
The argument relies on the computation of the expected maximum of Brownian motion
under stochastic resetting~\cite{mmss}, together with a scaling analysis of the formula
for the generating series of the mean number of records $\mean{R_n}$ given in~\cite{m2s2}
(which coincides with~(\ref{eq32})).
Furthermore, it is conjectured in~\cite{m2s2} that~(\ref{mrequiv})
is still valid for an arbitrary value of the resetting probability $0<r<1$.
However, as explained above and further demonstrated in section~\ref{gumregime},
this conjecture does not hold true.

\subsection{A key scaling result}
\label{scakey}

The main purpose of this section is the derivation of the scaling
formula~(\ref{key})
which encodes the asymptotic distribution of the maximum~$M_n$
throughout the diffusive scaling regime
of long walks ($n\gg1$) in the presence of a weak resetting ($r\ll1$),
for an arbitrary continuous symmetric step length distribution with finite variance.
The asymptotic equivalence~(\ref{mrequiv}) ensures that the same result also
describes the statistics of the number $R_n$ of records in the same regime.
The scaling formula~(\ref{key}) was mentioned in~\cite{em,mmss}, where
it has however not been exploited in detail.
Our present goal is to recover this result from the general formalism of section~\ref{gal},
and to analyze it at full length.
For this purpose, rather than using the general solution~(\ref{qfull}),
we find it more instructive to solve directly~(\ref{qeq}) within the diffusion approximation.
This amounts to approximating the integral involved in~(\ref{qeq}) as
\beqa
\fl
I&=&\int_{-\infty}^{M-x}\w Q(z,x+\eta,M)\rho(\eta)\dd\eta
\nonumber\\
\fl
&=&\int_{-\infty}^{M-x}
\left[\w Q(z,x,M)+\eta\w q'(z,x,M)+\frac{\eta^2}{2}\w
Q''(z,x,M)+\cdots\right]
\rho(\eta)\dd\eta
\nonumber\\
\fl
&=&\w Q(z,x,M)+D\w Q''(z,x,M)+\cdots,
\eeqa
where accents denote derivatives with respect to~$x$.
Furthermore,
setting
\beq
z=\e^{-s},
\eeq
the diffusive scaling regime corresponds to $s$ and $r$ being simultaneously small.
In this regime, the integral equation~(\ref{qeq}) comes down to
\beq
-D\w Q''(z,x,M)+(r+s)\w Q(z,x,M)\approx1+r\w Q(z,0,M),
\eeq
with boundary condition $\w Q(z,M,M)=0$.
The solution which remains bounded as $x\to-\infty$ reads
\beq
\w Q(z,x,M)\approx\frac{1-\e^{\mu(x-M)}}{s+r\,\e^{-\mu M}},
\label{qres}
\eeq
with
\beq
\mu=\sqrt\frac{s+r}{D}.
\label{mures}
\eeq
The expression~(\ref{qres}) has the structure of~(\ref{qfull}),
as could be expected, with
\beq
\w q(z,y)\approx\frac{1-\e^{-y\sqrt{s/D}}}{s}.
\label{gdiff}
\eeq
In the diffusive scaling regime,
generating series become Laplace transforms with respect to $n$, with conjugate variable $s$.
The expressions~(\ref{fqser}) and~(\ref{qres}) translate to
\beq
\lap{n} F_n(M)\approx\frac{1-\e^{-\mu M}}{s+r\,\e^{-\mu M}}.
\label{key}
\eeq

In the remainder of section~\ref{sca}
we analyze the consequences of~(\ref{key})
throughout the diffusive scaling regime, in the presence of a weak resetting.
For now, prior to this, let us discuss the situation without resetting.
For $r=0$,~(\ref{key}) reads
\beq
\lap{n} F_n(M)\approx\frac{1-\e^{-M\sqrt{s/D}}}{s}
\eeq
(see~(\ref{fqzero}),~(\ref{gdiff})).
Performing the inverse Laplace transform, we obtain
\beq
F_n(M)\approx\erf\frac{M}{2\sqrt{Dn}},
\eeq
where erf is the error function,
thus
\beq
f_n(M)\approx\frac{\e^{-M^2/(4Dn)}}{\sqrt{\pi Dn}}.
\label{mgau}
\eeq
Using the asymptotic equivalence~(\ref{mrequiv}),
this translates to
\beq
p_n(R)\approx\frac{\e^{-R^2/(4n)}}{\sqrt{\pi n}}.
\label{rgau}
\eeq
The corresponding mean values scale as
\beq
\mean{M_n}\approx2\sqrt\frac{Dn}{\pi},\qquad
\mean{R_n}\approx2\sqrt\frac{n}{\pi}.
\label{mrmoms}
\eeq

We have thus recovered the well-known facts
that the asymptotic distributions of~$M_n$ and~$R_n$
are half-Gaussians for all continuous step length distributions with finite variance.
In the case of the maximum, this asymptotic law is related to a known property of Brownian motion,
namely that the maximum and the absolute value of the current position
have the same distribution.
This comes as a consequence of the reflection principle~\cite{bachelier,levy,borodin}.
In the case of the number of records,
the half-Gaussian asymptotic law, and in fact the full statistics of $R_n$ for finite $n$,
are universal among random walks with a continuous symmetric step length distribution
and finite or infinite variance~\cite{feller2,ziff} (see also the review~\cite{revue}
and~section~\ref{wo} for details).

\subsection{Moments of maximum and number of records}
\label{scamoms}

We now analyze the consequences of~(\ref{key})
for the behaviour of the moments of the maximum $M_n$
and, using~(\ref{mrequiv}), of the number $R_n$ of records in the diffusive scaling regime, in the presence of a weak resetting.
For all integers $k\ge1$, we have
\beq
\mean{M_n^k}=\int_0^\infty M^k\,f_n(M)\dd M
=k\int_0^\infty M^{k-1}(1-F_n(M))\dd M.
\label{momparts}
\eeq
In Laplace space, this reads
\beqa
\lap{n}\mean{M_n^k}
&=&k\int_0^\infty M^{k-1}\left(\frac{1}{s}-\lap{n} F_n(M)\right)\dd M
\nonumber\\
&\approx&\frac{k(s+r)}{s}\int_0^\infty\frac{M^{k-1}}{r+s\,\e^{\mu M}}\,\dd M
\nonumber\\
&\approx&\frac{kD^{k/2}}{s(s+r)^{k/2-1}}\int_0^\infty\frac{p^{k-1}}{r+s\,\e^p}\,\dd p.
\label{momslap}
\eeqa
The second line ensues from~(\ref{key}),
and the third one from setting $p=\mu M$ (see~(\ref{mures})).
The latter expression is a homogeneous function of the variables $s$ and $r$
with degree $d=-k/2-1$.
This implies the scaling behavior
\beq
\mean{M_n^k}\approx(Dn)^{k/2}\,\phi_k(u),\qquad
\mean{R_n^k}\approx n^{k/2}\,\phi_k(u),
\label{momsca}
\eeq
where the scaling variable $u=nr$
is the mean number of resettings~(see~(\ref{udef})).

The scaling functions $\phi_k(u)$ can be derived by performing the inverse
Laplace transform of~(\ref{momslap}).
Introducing the ratio
\beq
\lam=\frac{s}{r},
\label{ldef}
\eeq
so that $ns=\lam u$,
we obtain
\beq
\phi_k(u)=ku^{-k/2}\int\frac{\dd\lam}{2\pi\ii}\frac{\e^{\lam
u}}{\lam(\lam+1)^{k/2-1}}\,L_k(\lam),
\label{philap}
\eeq
with
\beqa
L_k(\lam)
&=&\int_0^\infty\frac{p^{k-1}}{1+\lam\,\e^p}\,\dd p
\nonumber\\
&=&-(k-1)!\sum_{m\ge1}\frac{1}{m^k}\!\left(-\frac{1}{\lam}\right)^m
\nonumber\\
&=&-(k-1)!\,\Li_k\!\left(-\frac{1}{\lam}\right),
\label{lkdef}
\eeqa
where $\Li_k$ are the polylogarithms.

The only case where the function $L_k(\lam)$ is elementary is $k=1$, where we
have
\beq
L_1(\lam)=\ln\frac{\lam+1}{\lam},
\eeq
hence
\beq
\phi_1(u)=\frac{1}{\sqrt{u}}\int\frac{\dd\lam}{2\pi\ii}\,\e^{\lam
u}\,\frac{\sqrt{\lam+1}}{\lam}
\ln\frac{\lam+1}{\lam}.
\label{phi1lap}
\eeq
The expression~(\ref{philap}) also somehow simplifies for $k=2$,
as it does not involve any power of $(\lam+1)$.
Using the power-series expansion~(\ref{lkdef}) of the dilogarithm, we obtain
\beq
\phi_2(u)=2\sum_{m\ge1}\frac{(-u)^{m-1}}{m^2\,m!}
=\frac{2}{u}\int_0^u\frac{1-\e^{-v}}{v}\ln\frac{u}{v}\,\dd v.
\eeq

The scaling functions $\phi_1(u)$ and $\phi_2(u)$,
corresponding to the first two moments, have been studied in~\cite{m2s2,em,mmss}.
An expression equivalent to~(\ref{philap}) for all scaling functions $\phi_k(u)$
has been derived recently,
by considering the maximum of Brownian motion with resetting~\cite{singh}.
All the scaling functions $\phi_k(u)$ are decreasing functions of $u=nr$.
In other words, at least in the scaling regime,
all integer moments of~$M_n$ and $R_n$ are maximal in the absence of resetting.
These scaling functions can be investigated analytically for small and large
values of $u$.

\subsubsection*{Behavior for $u\ll1$.}

The behavior of $\phi_k(u)$ for small $u$ can be derived
by expanding the integrand of~(\ref{philap}) as an inverse power series for
large $\lam$.
For generic values of the integer order $k$,
keeping only the two leading terms, we obtain
\beqa
\phi_k(u)
&=&k!\,u^{-k/2}\int\frac{\dd\lam}{2\pi\ii}\frac{\e^{\lam u}}{\lam^{k/2+1}}
\left[1+\left(1-\frac{k}{2}-\frac{1}{2^k}\right)\frac{1}{\lam}+\cdots\right]
\nonumber\\
&=&\frac{2^k}{\sqrt\pi}\,\Gamma\!\left(\frac{k+1}{2}\right)
\left(1-\frac{k-2+2^{1-k}}{k+2}\,u+\cdots\right).
\label{phikser}
\eeqa
The values of $\phi_k(0)$ are the rescaled moments of the half-Gaussian
laws~(\ref{mgau}),~(\ref{rgau}),
which hold in the absence of resetting.
The first correction term, proportional to $u$, vanishes for $k=1$,
whereas it is negative for all higher values of $k$.

For $k=1$,
we can obtain a few more terms by
expanding the integrand of~(\ref{phi1lap}) for large $\lam$:
\beq
\phi_1(u)=\frac{2}{\sqrt\pi}\left(1-\frac{u^2}{90}+\frac{u^3}{315}+\cdots\right).
\label{phi1ser}
\eeq

\subsubsection*{Behavior for $u\gg1$.}
\label{ularge}

The behavior of $\phi_k(u)$ for large $u$ can be derived
by estimating the expression~(\ref{philap}) for small $\lam$ as follows:
\beqa
\phi_k(u)
&\approx&ku^{-k/2}\int\frac{\dd\lam}{2\pi\ii}\frac{\e^{\lam u}}{\lam}
\int_0^\infty\frac{p^{k-1}\,\dd p}{1+\lam\,\e^p}
\nonumber\\
&\approx&ku^{-k/2}\int_0^\infty p^{k-1}\,\dd p
\int\frac{\dd\lam}{2\pi\ii}\frac{\e^{\lam u}}{\lam(1+\lam\,\e^p)}
\nonumber\\
&\approx&ku^{-k/2}\int_0^\infty p^{k-1}(1-\exp(-u\,\e^{-p}))\dd p
\nonumber\\
&\approx&u^{1-k/2}\int_0^\infty p^k\,\exp(-u\,\e^{-p})\dd p.
\label{phitogum}
\eeqa
The third line is obtained by integrating over $\lam$,
and the fourth one by an integration by parts.
Setting
\beq
p=\ln u+\xi,
\eeq
the expression~(\ref{phitogum}) becomes
\beq
\phi_k(u)\approx u^{-k/2}\int_{-\infty}^\infty(\ln u+\xi)^k
\stackunder{\underbrace{\exp(-\xi-\e^{-\xi})\dd\xi}}{\rm Gumbel}.
\label{gumbel}
\eeq
We recognize the density of the canonical Gumbel random variable $G$, such that
\beq\label{loigumbel}
\prob(\xi<G<\xi+\dd\xi)=\exp(-\xi-\e^{-\xi})\dd\xi.
\eeq
The expression~(\ref{gumbel}) therefore implies that, for large $u$, $M_n$ and $R_n$ behave as
\beq
M_n\approx\sqrt\frac{D}{r}\,(\ln u+G),\qquad
R_n\approx\frac{\ln u+G}{\sqrt{r}}.
\label{mrgum}
\eeq
These estimates involve the sum of a large deterministic term $\ln u$
and a fluctuating term $G$ of order unity,
either positive or negative, distributed according to the Gumbel law~(\ref{loigumbel}).
We have in particular $\mean{G}=\gamma$, where $\gamma$ is Euler's constant, and so
\beq
\phi_1(u)\approx\frac{\ln u+\gamma}{\sqrt{u}},
\eeq
so that the mean values of $M_n$ and $R_n$ read
\beq
\mean{M_n}\approx\sqrt\frac{D}{r}\,(\ln nr+\gamma),\qquad
\mean{R_n}\approx\frac{\ln nr+\gamma}{\sqrt{r}}.
\label{meangum}
\eeq

In order to characterize the fluctuations of $M_n$ and $R_n$ around the above mean value,
it is convenient to estimate their cumulants.
Equation~(\ref{mrgum}) yields
\beq
\cum{M_n^k}\approx\left(\frac{D}{r}\right)^{k/2}c_k,\qquad
\cum{R_n^k}\approx\frac{c_k}{r^{k/2}}\qquad(k\ge2),
\eeq
where the cumulants $c_k=\cum{G^k}$ are pure numbers, which can be evaluated
by means of
their generating series
\beq
\sum_{k\ge1}\frac{c_ky^k}{k!}=\ln\mean{\e^{yG}}=\ln\Gamma(1-y),
\eeq
hence
\beq
c_k=(k-1)!\zeta(k).
\eeq
Here, $\zeta(k)$ denotes the value of Riemann's zeta function at the integer $k\ge2$.
We have in particular $c_2=\pi^2/6$, hence the variances approach the finite
limits
\beq
\var M_n\approx\frac{\pi^2D}{6r},\qquad
\var R_n\approx\frac{\pi^2}{6r}.
\eeq
The first scaling functions read
\beqa
\phi_1(u)\approx\frac{L}{\sqrt{u}},
\nonumber\\
\phi_2(u)\approx\frac{1}{u}(L^2+c_2),
\nonumber\\
\phi_3(u)\approx\frac{1}{u^{3/2}}(L^3+3c_2L+c_3),
\nonumber\\
\phi_4(u)\approx\frac{1}{u^2}(L^4+6c_2L^2+4c_3L+c_4+3c_2^2),
\eeqa
and so on, with the shorthand notation
\beq
L=\ln u+\gamma.
\eeq
The corrections to the above estimates are of relative order $1/u$, up to logarithms
(see~(\ref{gumcor})), i.e., exponentially small in $L$.

The logarithmic behavior~(\ref{meangum}) has been interpreted
in~\cite{m2s2,em,mmss} in terms of extreme-value statistics.
The expression~(\ref{mrgum}) corroborates this interpretation.
When the scaling variable $u$---the mean number of resetting events---is large,
the actual number of resettings up to time $n$ is close to $u$.
The maximum $M_n$ is therefore approximately equal to the maximum of $u$
iid random variables $m_i$, each of them being the maximum of the walker's position
in a stretch of random walk between two successive resetting events.
For $r\ll1$, the distribution of each local maximum $m_i$ reads approximately
\beq
p(m)\approx\mu_0\,\e^{-\mu_0 m},
\label{fm}
\eeq
with $\mu_0=\sqrt{r/D}$.
The distribution~(\ref{fm}) can be obtained by averaging
the half-Gaussian distribution~(\ref{mgau}) of~$M_n$
over the broad geometric distribution~(\ref{geolaw})
of the lapses of time between successive resetting events.
The maximum of a large number~$u$ of iid random variables distributed according
to the distribution~(\ref{fm}) is precisely given by~(\ref{mrgum}).

\subsection{Distributions of the maximum and of the number of records}
\label{scafull}

Let us now investigate the scaling form of the distribution
of the maximum~$M_n$, or equivalently, of the number $R_n$ of records in the diffusive scaling regime.

The scaling behavior~(\ref{momsca}) of the moments translates to the following
scaling laws for the distributions defined in~(\ref{fdef}),~(\ref{pdef}):
\beq
f_n(M)\approx\frac{1}{\sqrt{Dn}}\,\Phi(X,u),\qquad
p_n(R)\approx\frac{1}{\sqrt{n}}\,\Phi(X,u),
\label{fsca}
\eeq
where the reduced distribution $\Phi(X,u)$ is a scaling function of the variables
\beq
X=\frac{M}{\sqrt{Dn}}\quad\hbox{or}\quad X=\frac{R}{\sqrt{n}},\qquad u=nr.
\label{vars}
\eeq

An explicit expression of $\Phi(X,u)$ can be derived from~(\ref{key}).
Differentiating the latter equation with respect to~$M$ yields
\beq
f_n(M)\approx\int\frac{\dd s}{2\pi\ii}\,
(s+r)^{3/2}\,\frac{\e^{ns-M\sqrt{s+r}}}{(s+r\,\e^{-M\sqrt{s+r}})^2}.
\eeq
Introducing the variables $X$ and $u$, as well as the ratio $\lam$ (see~(\ref{ldef})),
we obtain
\beq
\Phi(X,u)=\sqrt{u}\int\frac{\dd\lam}{2\pi\ii}\,
(\lam+1)^{3/2}\,\frac{\e^{\lam u-wX}}{(\lam+\e^{-wX})^2},
\label{phidef}
\eeq
with the shorthand notation
\beq
w=\sqrt{u(\lam+1)}.
\eeq
It can be checked that the reduced distribution $\Phi(X,u)$ obeys the sum rules
\beqa
\int_0^\infty\Phi(X,u)\dd X=1,
\nonumber\\
\int_0^\infty X^k\,\Phi(X,u)\dd X=\phi_k(u)\qquad(k\ge1),
\eeqa
as should be, where the functions $\phi_k(u)$ are given by~(\ref{philap}).

\begin{figure}[!ht]
\begin{center}
\includegraphics[angle=0,width=.7\linewidth,clip=true]{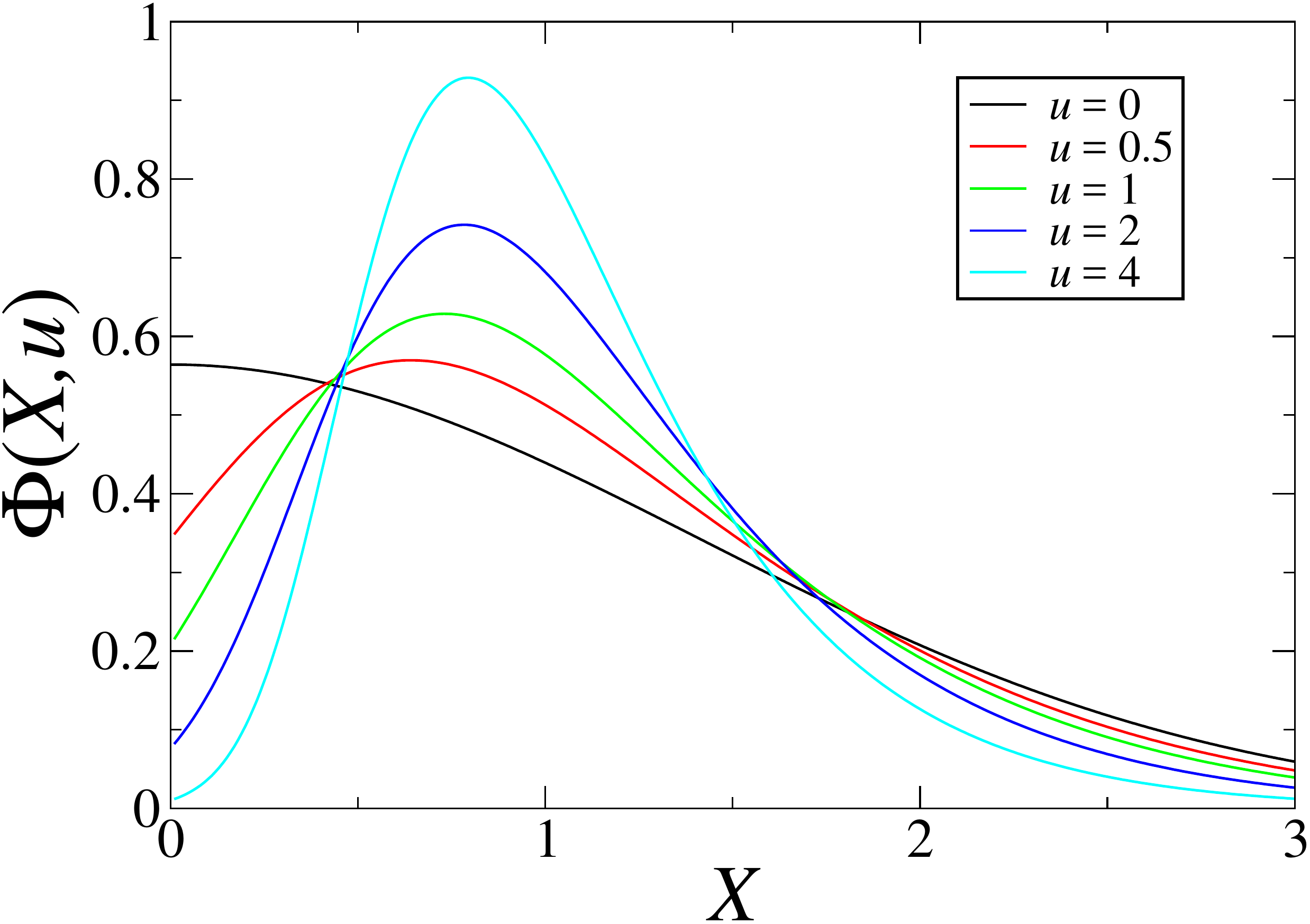}
\caption{\small
Reduced distribution $\Phi(X,u)$ against $X$ for several values of $u$ (see legend).}
\label{Phi}
\end{center}
\end{figure}

As the mean number $u=nr$ of resettings increases,
the reduced distribution $\Phi(X,u)$ interpolates between a half-Gaussian law at $u=0$
(see~(\ref{mgau}),~(\ref{rgau}),~(\ref{smallu}))
and a Gumbel law at $u\gg1$ (see~(\ref{mrgum}),~(\ref{gumcor})).
This is illustrated in figure~\ref{Phi},
showing $\Phi(X,u)$ against $X$ for several values of $u$.
The data have been obtained by means of a numerical evaluation
of the contour integral in~(\ref{phidef}).

At any fixed value of $u$,
there is a most probable value $X_0(u)$,
where the reduced distribution $\Phi(X,u)$ is maximal.
For $u\ll1$, we have
\beq
X_0(u)\approx4\sqrt{\pi}\,u,
\eeq
as a consequence of the expansion~(\ref{smallx}).
For $u\gg1$, we have
\beq
X_0(u)\approx\frac{\ln u}{\sqrt{u}},
\eeq
since the most probable value of the Gumbel variable $G$ is zero
(see~(\ref{xtoxi}),~(\ref{largeu})).
The most probable value $X_0(u)$ thus tends to zero both at small $u$ and at
large~$u$.
It is maximal at an intermediate value of $u$,
namely $u\approx3.4111$,
where it equals $X_0\approx0.795837$.

\begin{figure}[!ht]
\begin{center}
\includegraphics[angle=0,width=.7\linewidth,clip=true]{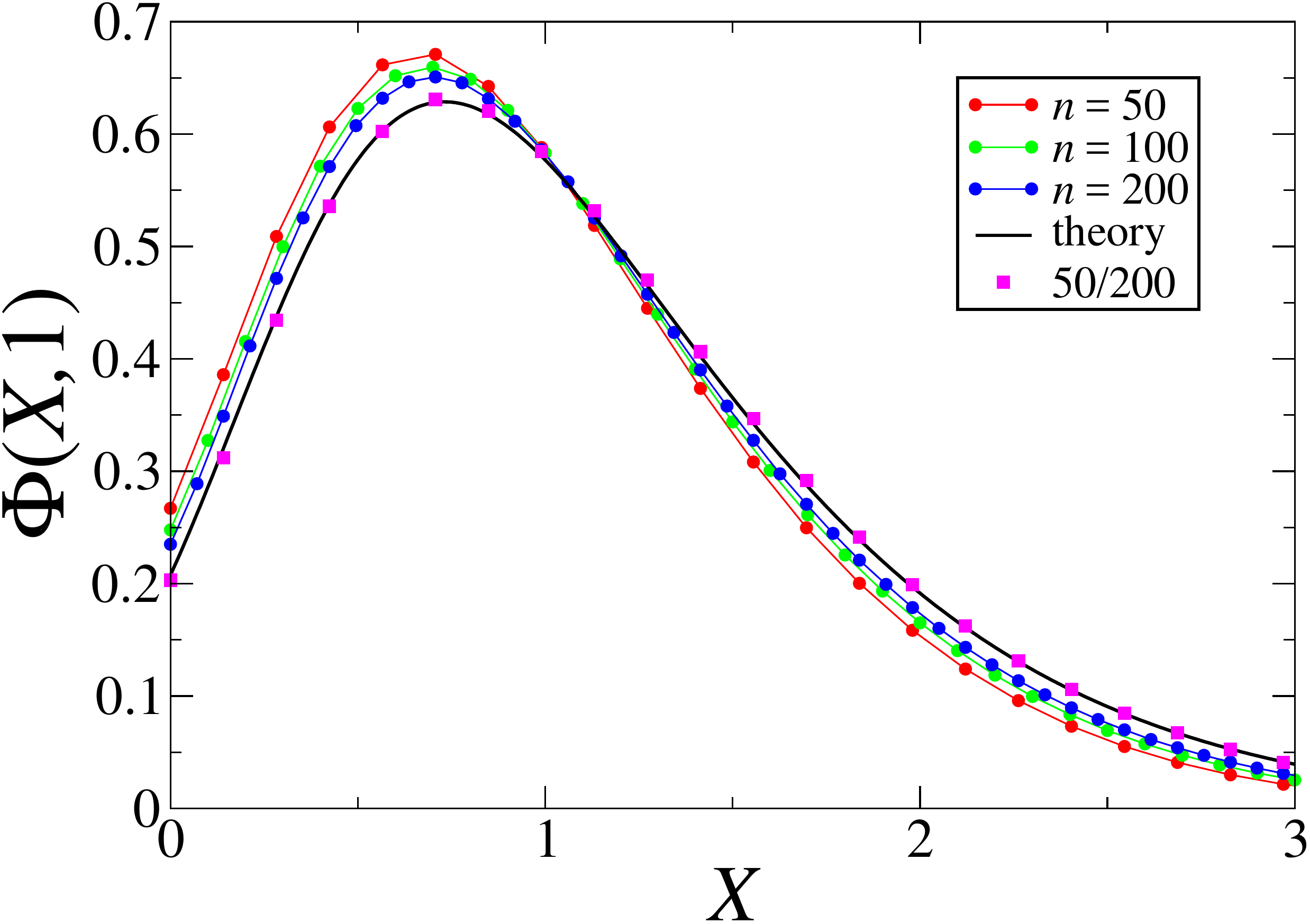}
\caption{\small
Full curves and symbols:
distribution of the number of records~$R_n$ of walks with uniformly
distributed step lengths, for $u=nr=1$ and several $n$ (see legend),
rescaled according to~(\ref{fsca}),~(\ref{vars}).
Black curve (theory): theoretical prediction~(\ref{phidef}).
Square symbols (50/200):
extrapolated values based on data with $n=50$ and $n=200$ (see text).}
\label{un}
\end{center}
\end{figure}

A comparison between the actual distribution of the number $R_n$ of records
of walks with a uniform step length distribution on the interval $[-1,+1]$,
measured by means of numerical simulations for $10^8$ walks of 50, 100, and 200 steps,
and the theoretical prediction~(\ref{phidef})
is shown in figure~\ref{un}.
The mean number of resettings is fixed to $u=nr=1$,
i.e., somewhat half way between the Gaussian and Gumbel limits.
The data points converge smoothly to the theoretical prediction~(\ref{phidef})
(black curve),
already shown in green in figure~\ref{Phi}.
The square symbols show extrapolated values based on data for 50 and 200 steps,
assuming that the leading correction to scaling is of relative order $1/\sqrt{n}$
(see e.g.~(\ref{mcorr}),~(\ref{mrrelat}) for examples).
The combination $2\sqrt{200}\,p_{200}(2R)-\sqrt{50}\,p_{50}(R)$
is tailored in order to eliminate the above leading correction,
giving thus a better estimate of $\Phi(X,u)$.
This combination is plotted against $X=R/\sqrt{50}$ for $R=0,1,\dots,21$.
These extrapolated data points are hardly distinguishable from the theoretical prediction,
thus providing a strong corroboration of the whole analysis.

To close this section,
we show how the behavior of the reduced distribution $\Phi(X,u)$
at small and large values of each of its arguments
can be studied analytically.

\subsubsection*{Behavior for $u\ll1$.}

The behavior of $\Phi(X,u)$ for small $u$ can be derived
by setting $\lambda=p^2/u$ in~(\ref{phidef})
and expanding the integrand as a power series in~$u$ at fixed $p$.
We thus obtain
\beqa
\Phi(X,u)
&=&\int\frac{\dd p}{2\pi\ii}\,\e^{p^2-pX}
\left[2+\left(\frac{3}{p^2}-\frac{X}{p}-\frac{4\e^{-pX}}{p^2}\right)u+\cdots\right]
\nonumber\\
&=&\frac{\e^{-X^2/4}}{\sqrt\pi}
\label{smallu}
\\
&+&
\left(\frac{3\e^{-X^2/4}-4\e^{-X^2}}{\sqrt\pi}+4X\erfc
X-2X\erfc\frac{X}{2}\right)u+\cdots,
\nonumber
\eeqa
where erfc is the complementary error function.
The first term reproduces the asymptotic half-Gaussian
distributions~(\ref{mgau}),~(\ref{rgau})
in the absence of resetting.

\subsubsection*{Behavior for $u\gg1$.}

The leading-order behavior of $\Phi(X,u)$ for large $u$ can be derived
by estimating the expression~(\ref{phidef}) for small $\lam$ as follows:
\beqa
\Phi(X,u)
&\approx&\sqrt{u}\int\frac{\dd\lam}{2\pi\ii}\,
\frac{\e^{\lam u-X\sqrt{u}}}{(\lam+\e^{-X\sqrt{u}})^2}
\nonumber\\
&\approx&u^{3/2}\exp\left(-X\sqrt{u}-u\e^{-X\sqrt{u}}\right).
\label{phigum}
\eeqa
Setting, in agreement with~(\ref{mrgum}),
\beq
X=\frac{\ln u+\xi}{\sqrt{u}},
\label{xtoxi}
\eeq
the estimate~(\ref{phigum}) translates to
\beq
\Phi(X,u)\dd X\approx\exp\left(-\xi-\e^{-\xi}\right)\dd\xi.
\label{largeu}
\eeq
The emergence of the Gumbel distribution for the fluctuating part $G$ introduced in~(\ref{mrgum})
is thus confirmed by the analysis of the distribution of $M_n$.
Its interpretation in terms of extreme-value statistics
was given at the end of section~\ref{ularge}.

The corrections to the above leading-order behavior
can be derived by taking higher powers of $\lam$ into account.
Skipping details, we only give the outcome
in the bulk of the distribution, i.e., for large $u$ at fixed~$\xi$:
\beqa
\Phi(X,u)
&\approx&\sqrt{u}\,\exp\left(-\xi-\e^{-\xi}\right)
\label{gumcor}
\\
&\times&\left[1+\frac{\ln
u+\xi}{2u}\,(3\e^{-\xi}-\e^{-2\xi}-1)+\frac{3(1-\e^{-\xi})}{2u}+\cdots\right].
\nonumber
\eeqa

\subsubsection*{Behavior for $X\ll1$.}

The behavior of $\Phi(X,u)$ for small $X$
can be derived by setting $\lam=p-1$ and expanding the integrand
of~(\ref{phidef}) as a power series in $X$.
We thus obtain
\beq
\Phi(X,u)=\e^{-u}
\left[\frac{1}{\sqrt\pi}+2uX+\left(6u^2-3u-\frac14\right)\frac{X^2}{\sqrt\pi}+\cdots\right].
\label{smallx}
\eeq
We have in particular
\beq
\Phi(0,u)=\frac{\e^{-u}}{\sqrt\pi}.
\label{phizero}
\eeq

\subsubsection*{Behavior for $X\gg1$.}

The behavior of $\Phi(X,u)$ for large $X$
can be derived by approximating~(\ref{phidef}) as
\beq
\Phi(X,u)\approx\sqrt{u}\int\frac{\dd\lam}{2\pi\ii}\,
\frac{\e^{\lam u-X\sqrt{u(\lam+1)}}}{\sqrt\lam},
\eeq
and evaluating the integral by the saddle-point method.
The saddle point sits~at
\beq
\lam_c=\frac{X^2}{4u}-1.
\label{lamc}
\eeq
We thus obtain
\beq
\Phi(X,u)\approx\frac{\e^{-u-X^2/4}}{\sqrt\pi}.
\label{philarge}
\eeq
For large $X$, the tail of the half-Gaussian law which prevails in the absence of resetting
therefore survives for all values of $u$.
Furthermore,~(\ref{phizero}) and~(\ref{philarge})
have the same exponential dependence in $u$.

If $u$ also becomes large,
the asymptotic law~(\ref{philarge}) still holds,
albeit with an $X$-dependent prefactor, as long as $\lam_c$ is positive, i.e.,
for $X>2\sqrt{u}$.
For $X\approx2\sqrt{u}$, $\Phi(X,u)$ exhibits a sharp crossover,
over a finite range of values of $X$,
between the exponential tail of the Gumbel law~(\ref{phigum}) and the Gaussian one~(\ref{philarge}).

\section{Beyond the diffusive scaling regime}
\label{beyond}

We hereafter illustrate on a few specific situations
to what extent the statistics of~$M_n$ and of $R_n$ may differ from each other
whenever the asymptotic equivalence~(\ref{mrequiv}) does not hold,
i.e., outside the diffusive scaling regime analysed in section~\ref{sca}.
This section is devoted to generic continuous and symmetric step length distributions,
whereas section~\ref{special} is devoted to two examples of distributions
for which the integral equation~(\ref{qeq}) can be solved by elementary means.

\subsection{Corrections to asymptotic behavior without resetting}
\label{wo}

We start with the situation in the absence of resetting.
In this case, the asymptotic results~(\ref{mgau}),~(\ref{rgau})
and~(\ref{mrmoms})
can be compared to more detailed results.

We start by recalling that the renewal structure of the record process
allows a simple derivation of the expression of the distribution of the number $R_n$ of records.
The corresponding generating function reads~\cite{revue,ziff}
(see~\cite[Sec.~3]{glrenew} for a short proof valid for any renewal process)
\beq
\sum_{n\ge0}p_n(R)z^n=\w q(z)\,\w f(z)^R,
\label{psa}
\eeq
where $\w q(z)$ and $\w f(z)$ are respectively given in~(\ref{qzres}) and~(\ref{fzres}).
Introducing the notation
\beq
\nu_0=\sqrt{1-z},
\eeq
we have $\w q(z)=1/\nu_0$ and $\w f(z)=1-\nu_0$, and so
\beq
\sum_{n\ge0}p_n(R)z^n=\frac{(1-\nu_0)^R}{\nu_0}.
\label{psae}
\eeq
As a consequence, we have
\beq\label{pnR}
p_n(R)=
\frac{(2n-R)!}{2^{2n-R}n!(n-R)!}=\frac{{{2n-R}\choose n}}{2^{2n-R}}\qquad(R=0,\dots,n).
\eeq
This distribution is universal, i.e., independent of the step length distribution,
whenever it is continuous and symmetric, either with a finite variance or not.

The generating series of the mean number of records evaluates to
\beq
\sum_{n\ge0}\mean{R_n}z^n
=\frac{1-\nu_0}{\nu_0^3}=\frac{1}{(1-z)^{3/2}}-\frac{1}{1-z},
\label{avere}
\eeq
hence
\beq
\mean{R_n}=(2n+1)b_n-1=2\sqrt\frac{n}{\pi}\left(1+\frac{3}{8n}-\frac{7}{128n^2}+\cdots\right)-1,
\label{rcorr}
\eeq
where $b_n$ is defined in~(\ref{bdef}).

The mean value $\mean{M_n}$ of the maximum of diffusive random walks
has been investigated in~\cite{cm,mcz}.
The quantity $\gamma$ introduced in~\cite{cm} reads $\gamma=-\ell/\sqrt{2D}$,
thereby
\beq
\mean{M_n}=2\sqrt\frac{Dn}{\pi}-\ell+\cdots,
\label{mcorr}
\eeq
where $\ell$ is the extrapolation length (see~(\ref{extra})).

The leading terms of the expansions~(\ref{rcorr}) and~(\ref{mcorr}) agree
with~(\ref{mrmoms})
and correspond to the asymptotic half-Gaussian
distributions~(\ref{mgau}),~(\ref{rgau})
of $M_n$ and~$R_n$.
Their first correction terms however differ.
The whole series of corrections in~(\ref{rcorr}) is universal,
i.e., independent of the step length distribution,
provided it is symmetric and continuous,
whereas the first correction in~(\ref{mcorr}),
involving the extrapolation length~$\ell$,
depends on the underlying distribution.
As a consequence, the first correction to the equivalence~(\ref{mrequiv}) for
generic diffusive walks
appears as a finite limit for the difference
\beq
\lim_{n\to\infty}\bigl(\mean{M_n}-\sqrt{D}\mean{R_n}\bigr)=\sqrt{D}-\ell.
\label{asydiff}
\eeq
This quantity vanishes for the symmetric exponential distribution considered in
section~\ref{se}.
In general it may be either positive or negative.

\subsection{Asymptotic behavior at finite resetting probability}
\label{gumregime}

We now consider long random walks with an arbitrary resetting probability $r$,
whereas the diffusive scaling regime studied in section~\ref{sca} corresponds to $r\ll1$.

The expression of the mean value $\mean{R_n}$ of the record number has been derived
in~\cite{m2s2} in full generality.
In the regime of current interest ($n\to\infty$, $r$ finite), it obeys the logarithmic law
\beq
\mean{R_n}\approx\frac{\ln(nr(1-\sqrt{r}))+\gamma}{\sqrt{r}}.
\label{rave}
\eeq
This expression only differs from its counterpart~(\ref{meangum})
in the scaling regime ($r\ll1$) by the $r$-dependence of the finite part of the
logarithm.

As far as the maximum $M_n$ is concerned,
no analytical prediction is available in general beyond the diffusive scaling
regime (see however sections~\ref{se} and~\ref{simple} for particular examples).
Let us propose the following heuristic line of thought.
For any value of the resetting probability $r$,
the distribution of the lapses of time between successive resetting events
is given by the geometric law~(\ref{geolaw}).
As a consequence, the positions of the walker have exponentially decaying
connected correlations,
thereby the successive positions form a sequence of nearly iid random variables.
The maximum~$M_n$ is therefore expected to be approximately distributed
as the maximum of an extensive number $n_\eff\approx c\,n$ of iid variables
distributed according to the steady-state distribution~$f(x)$ described in~\ref{steady}.
The factor $c$ somehow takes
the above mentioned correlations into account in an effective way.
Its dependence on model parameters is not predicted by the present reasoning.
The dichotomy put forward in~\ref{steady} has the following consequences.

\begin{figure}[!ht]
\begin{center}
\includegraphics[angle=0,width=.7\linewidth,clip=true]{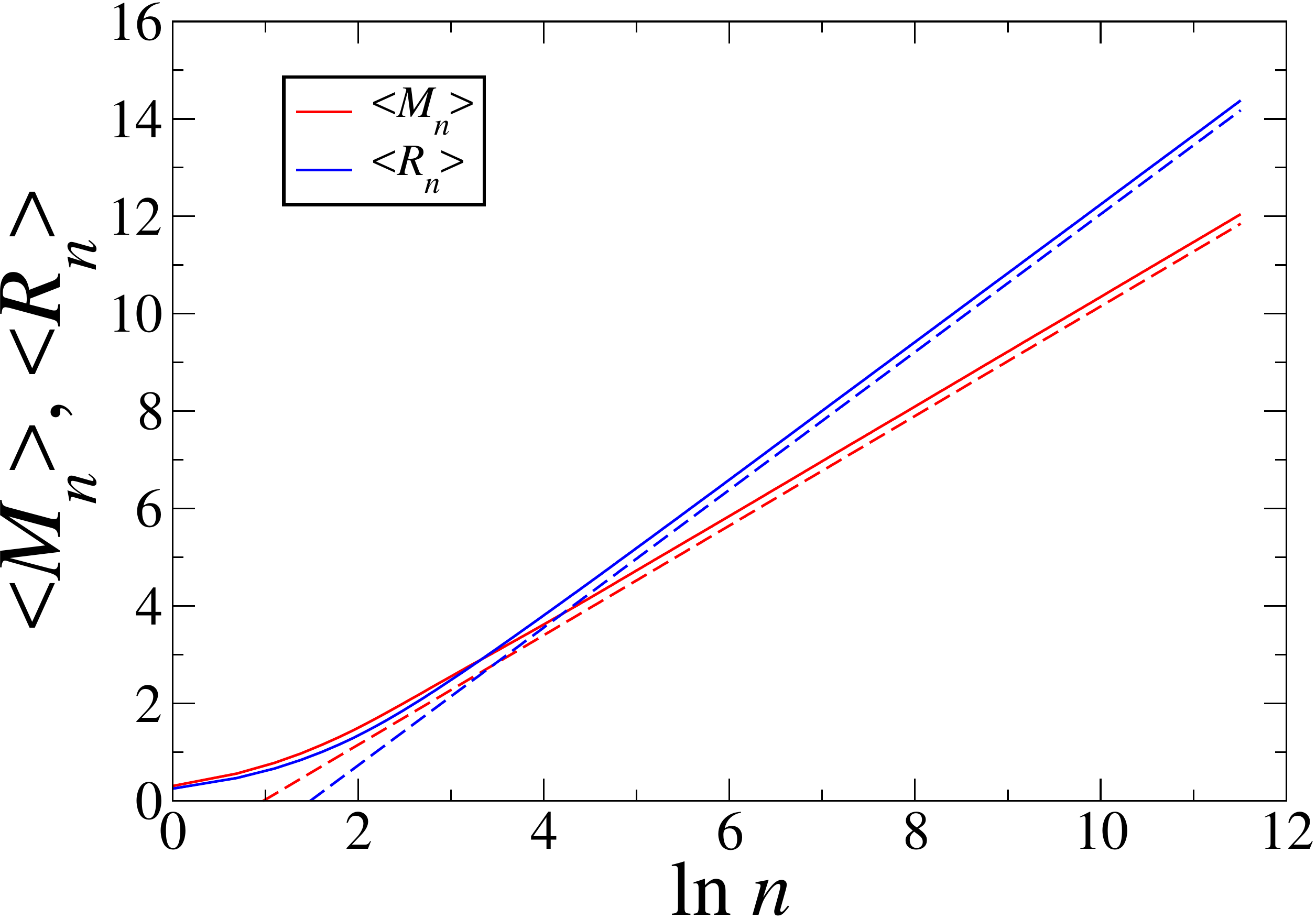}
\caption{\small
Full curves:
mean values $\mean{M_n}$ (red) and $\mean{R_n}$ (blue) against $\ln n$,
as measured by means of a numerical simulation up to $n=10^5$
for a uniform step length distribution on $[-w,w]$ with $w=\sqrt{6}$ and a resetting
probability $r=1/2$.
Dashed lines have slopes $1/\Kst\approx1.125002$ for $\mean{M_n}$
and $\sqrt{2}$ for $\mean{R_n}$.}
\label{aruni}
\end{center}
\end{figure}

For exponential and superexponential step length distributions,
the exponential tail~(\ref{fstatexp}) of the steady-state distribution $f(x)$
translates to
\beq
\mean{M_n}\approx\frac{\ln n}{\Kst}.
\label{mave}
\eeq
This estimate is robust, in the sense that it holds irrespective of the constant $c$.
In the diffusive scaling regime ($r\ll1$),
the decay rate $\Kst$ is given by~(\ref{ksca}),
with the consequence that the logarithmic growth laws~(\ref{rave})
and~(\ref{mave})
are related to each other according to the identity~(\ref{mrequiv}), as should be.
For an arbitrary resetting probability~$r$,
the decay rate $\Kst$ depends on $r$ and on details of the step length distribution.
Consider for definiteness a uniform distribution on the interval $[-w,w]$,
so that $\sigma^2=w^2/3$, $D=w^2/6$ and $\F\rho(k)=(\sin kw)/kw$.
The decay rate $\Kst$ is therefore given by the implicit equation
(see~(\ref{kdef}))
\beq
(1-r)\frac{\sinh\Kst w}{\Kst w}=1.
\label{kuni}
\eeq
Figure~\ref{aruni} shows the mean values $\mean{M_n}$ and $\mean{R_n}$ against $\ln n$,
as measured by means of a numerical simulation up to $n=10^5$
for a uniform step length distribution
and a resetting probability $r=1/2$.
We have chosen $w=\sqrt{6}$, hence $D=1$,
allowing a fair comparison between the two quantities.
Both datasets exhibit a logarithmic growth with the theoretically predicted
amplitudes (dashed lines),
namely $1/\Kst\approx1.125002$ for $\mean{M_n}$ and $1/\sqrt{r}=\sqrt{2}$ for $\mean{R_n}$.

For subexponential step length distributions,
the estimate~(\ref{fsub}) of the tails of the steady-state distribution
implies that $\mean{M_n}$ grows faster than a logarithm.
In other words, there is now a qualitative difference between the asymptotic
behavior
of $\mean{M_n}$ and of $\mean{R_n}$.
In the case where the step length distribution decays as a power law of the
form
\beq\label{eq:Lev}
\rho(\eta)\approx\frac{A}{\abs{\eta}^{1+\theta}}\qquad(\eta\to\pm\infty),
\eeq
with an arbitrary tail exponent $\theta>1$,
the above heuristic reasoning implies
\beq
\mean{M_n}\approx B\,n^{1/\theta}.
\label{maxth}
\eeq
The mean maximum now grows as a power of the number of steps $n$,
in strong contrast with the universal logarithmic growth~(\ref{rave}) of the mean record number.
The prefactor~$B$ is not predicted here, as it depends on the unknown constant $c$.
For diffusive walks ($\theta>2$),
$\mean{M_n}$ grows with the exponent $1/\theta<1/2$ for $r>0$
and with the exponent $1/2$ for $r=0$.
The presence of resetting events therefore diminishes qualitatively the growth
of $\mean{M_n}$.
For L\'evy walks with $1<\theta<2$,~$\mean{M_n}$ grows with the exponent
$1/\theta$,
irrespective of the presence of resetting.
For L\'evy walks with $\theta<1$,
the mean absolute step length $\mean{\abs{\eta}}$ diverges,
and so does the mean maximum.
In spite of this,
the growth law~(\ref{maxth}) still holds for the typical value of~$M_n$,
again irrespective of the presence of resetting.
Finally, a crossover to the logarithmic prediction~(\ref{meangum}) is expected at small $r$ in all cases.

\begin{figure}[!ht]
\begin{center}
\includegraphics[angle=0,width=.7\linewidth,clip=true]{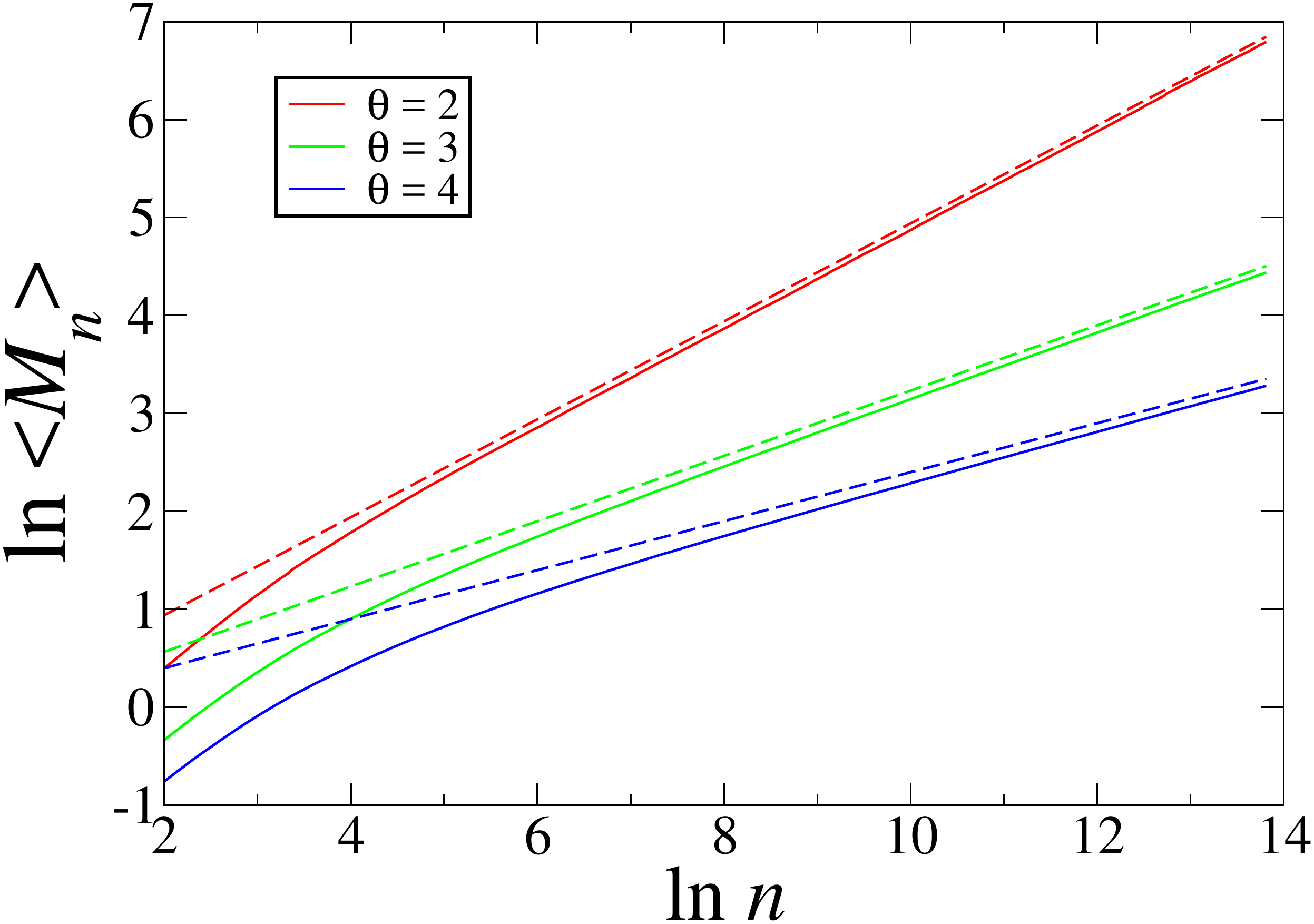}
\caption{\small
Full curves:
log-log plot of $\mean{M_n}$ against $n$
for the step length distribution~(\ref{rhoth}) with $\theta=2$, 3 and 4 (see legend)
and a resetting probability $r=1/2$.
Dashed lines have slopes 1/2, 1/3 and 1/4.}
\label{artheta}
\end{center}
\end{figure}

In order to check the validity of the scaling law~(\ref{maxth}),
we have measured by means of a numerical simulation the mean record number
$\mean{M_n}$
up to $n=10^5$ for step lengths of the form
$\eta_n=\eps_n(u_n^{-1/\theta}-1)$, where $\eps_n=\pm1$ with equal
probabilities
and~$u_n$ are uniform random variables over $[0, 1]$, resulting in
\beq
\rho(\eta)=\frac{\theta}{2(\abs{\eta}+1)^{1+\theta}}.
\label{rhoth}
\eeq
Figure~\ref{artheta} shows a log-log plot of $\mean{M_n}$ against $n$,
for the step length distribution~(\ref{rhoth}) with $\theta=2$, 3 and 4,
and a resetting probability $r=1/2$.
All datasets are in good agreement with the prediction~(\ref{maxth}) (dashed lines).

\section{Two particular examples}
\label{special}

The two particular examples discussed in this section have the virtue of
lending themselves to exact calculations at finite times.
The integral equation~(\ref{qeq}) can indeed be solved by elementary means.

\subsection{Symmetric exponential step length distribution}
\label{se}

We first address the case of the symmetric exponential distribution
(also known as the Laplace distribution)
\beq
\rho(\eta)=\frac{\e^{-\abs{\eta}}}{2},
\label{sedef}
\eeq
with variance $\sigma^2=2$ and diffusion coefficient $D=1$.

Let us first determine the steady-state distribution $f(x)$ of the walker's
position.
We have
\beq
\F\rho(k)=\frac{1}{1+k^2},
\eeq
thus, using~(\ref{fstatres}),
\beq
\F f(k)=\frac{r(1+k^2)}{r+k^2},
\eeq
and
\beq
f(x)=r\delta(x)+(1-r)\frac{\sqrt{r}}{2}\,\e^{-\sqrt{r}\abs{x}}.
\eeq
The decay rate of the continuous component therefore reads $\Kst=\sqrt{r}$
for all values of the resetting probability $r$.

The present situation is one of the rare examples
where the integral equation~(\ref{qeq}) can be solved by elementary means.
This equation has indeed an exact solution of the form $\w
Q(z,x,M)=A+B\e^{\nu x}$, that is,
\beq
\w Q(z,x,M)=\frac{1-(1-\nu)\e^{\nu(x-M)}}{1-z+rz(1-\nu)\e^{-\nu
M}}\qquad(x\le M),
\label{qse}
\eeq
with
\beq
\nu=\sqrt{1-\rz}
\label{nudef}
\eeq
(see~(\ref{rzdef})).
The expression~(\ref{qse}) again has the structure of~(\ref{qfull}),
as expected, with
\beq
\w q(z,y)=\frac{1-(1-\nu_0)\e^{-\nu_0 y}}{\nu_0^2},
\label{gexp}
\eeq
and
\beq
\nu_0=\sqrt{1-z}.
\eeq
The result~(\ref{gexp}) can be alternatively derived by means of the Pollaczek-Spitzer formula~(\ref{lres}).
We thus obtain the general formula (see~(\ref{fqser}) or~(\ref{fqser+}))
\beq
\w F(z,M)=\frac{1-(1-\nu)\e^{-\nu M}}{1-z+rz(1-\nu)\e^{-\nu M}},
\label{fse}
\eeq
encoding the exact distribution of the maximum $M_n$, for all values of $n$ and $r$.

In the scaling regime where $n$ is large and $r$ is small,
setting $z=\e^{-s}$ where $s$ is also small,
$\nu$ becomes $\mu$ (see~(\ref{mures})), and so~(\ref{fse})
becomes~(\ref{key}).
This provides a strong confirmation of the validity
of the diffusion approximation in the scaling regime.

At variance with the prediction~(\ref{qres}) of the diffusion approximation,
which vanishes at $x=M$ by construction,
the extrapolated exact solution~(\ref{qse}) vanishes at the point
\beq
x=M-\frac{\ln(1-\nu)}{\nu}=M+1+\frac{\nu}{2}+\cdots,
\eeq
slightly outside the domain $x\le M$.
The first correction term, which is the only one surviving in the scaling regime ($\nu\to0$),
yields the extrapolation length $\ell=1$.
This result can be recovered in two alternative ways,
either by a direct evaluation of the integrals entering~(\ref{extra}),
or by noting that the solution of the homogeneous
Milne equation~(\ref{milne}) is $H(y)=y+1$.

In order to pursue,
let us first consider the mean value $\mean{M_n}$ of the maximum in the presence of resetting.
The corresponding generating series reads
\beqa
\sum_{n\ge0}\mean{M_n}z^n
&=&\int_0^\infty\left(\frac{1}{1-z}-\w F(z,M)\right)\dd M
\nonumber\\
&=&\frac{\nu(1-\nu)}{1-z}\int_0^\infty\frac{1}{rz(1-\nu)+(1-z)\e^p}\,\dd p
\nonumber\\
&=&\frac{\nu}{rz(1-z)}\ln\frac{\nu(\nu-rz)}{1-z}.
\label{eq32}
\eeqa
The first line is obtained by using the identity~(\ref{momparts}) for $k=1$,
and the second one by using~(\ref{fse}) and setting $p=\nu M$.
The resulting expression (\ref{eq32}) coincides with the formula
for the generating series of the mean number of records $\mean{R_n}$ given
in~\cite{m2s2}.
This coincidence implies
\beq
\mean{M_n}=\mean{R_n}
\label{iden}
\eeq
identically for all $n$ and $r$.
In particular, the expression~(\ref{asydiff}) vanishes, since $\ell=D=1$.
The noteworthy identity~(\ref{iden}) contrasts with the generic case illustrated in figure~\ref{aruni}.
This identity cannot extend to the full distributions,
for the mere reason that $M_n$ is a real variable, while $R_n$ is an integer one.

The distributions of $M_n$ and $R_n$ can be compared to one another in more detail
in the absence of resetting, by considering their moments.
For $r=0$, the expression~(\ref{fse}) for $\w F(z,M)$
identifies to~(\ref{gexp}) (see~(\ref{fqzero})).
We have therefore
\beqa
\sm{k}(z)
&=&\sum_{n\ge0}\mean{M_n^k}z^n
\nonumber\\
&=&k\int_0^\infty M^{k-1}\left(\frac{1}{\nu_0^2}-\w F(z,M)\right)\dd M
\nonumber\\
&=&\frac{k(1-\nu_0)}{\nu_0^2}\int_0^\infty M^{k-1}\,\e^{-\nu_0 M}\,\dd M
\nonumber\\
&=&\frac{(1-\nu_0)k!}{\nu_0^{k+2}}.
\label{labs}
\eeqa
On the other hand, using~(\ref{psae}), we have
\beq
\sr{k}(z)
=\sum_{n\ge0}\mean{R_n^k}z^n
=\frac{1}{\nu_0}\sum_{R\ge0}R^k(1-\nu_0)^R.
\label{qabs}
\eeq
These series are studied in~\ref{stirling}.
In particular, it is shown
that the moments $\mean{M_n^k}$ and $\mean{R_n^k}$ obey the following linear identities
\beq
\mean{M_n^k}=\sum_{j=1}^k\sone{k}{j}\mean{R_n^j},\qquad
\mean{R_n^k}=\sum_{j=1}^k(-1)^{k-j}\stwo{k}{j}\mean{M_n^j},
\label{idens}
\eeq
where $\sone{k}{j}$ and $\stwo{k}{j}$ are respectively the Stirling numbers
of the first and of the second kind.
Besides~(\ref{iden}), the first few of these identities read
\beqa
\mean{M_n^2}=\mean{R_n^2}+\mean{R_n},
\nonumber\\
\mean{M_n^3}=\mean{R_n^3}+3\mean{R_n^2}+2\mean{R_n},
\nonumber\\
\mean{R_n^2}=\mean{M_n^2}-\mean{M_n},
\nonumber\\
\mean{R_n^3}=\mean{M_n^3}-3\mean{M_n^2}+\mean{M_n}.
\label{idenexs}
\eeqa

To leading order for large $n$, the two sequences of moments coincide,
in agreement with the analysis of the diffusive scaling regime.
The half-Gaussian laws~(\ref{mgau}),~(\ref{rgau}) imply (see~(\ref{phikser}))
\beq
\mean{M_n^k}\approx\mean{R_n^k}\approx
\frac{2^k}{\sqrt\pi}\,\Gamma\!\left(\frac{k+1}{2}\right)\,n^{k/2}.
\label{momlead}
\eeq

The identities~(\ref{idens}),~(\ref{idenexs}) show that,
except for the first one, all moments of $M_n$ are larger than those of~$R_n$.
The relative difference between $\mean{M_n^k}$ and $\mean{R_n^k}$
is however expected to become smaller and smaller for large $n$,
in order to conform with the analysis of the diffusive scaling regime.
This difference can be estimated from~(\ref{idens}), where the leading
correction corresponds to $j=k-1$.
Using $\sone{k}{k-1}=k(k-1)/2$, as well as the leading-order
result~(\ref{momlead}), we obtain
\beq
\mean{M_n^k}=\mean{R_n^k}\left(1+\frac{a_k}{\sqrt{n}}+\cdots\right),
\label{mrrelat}
\eeq
with
\beq
a_k=\frac{k(k-1)\Gamma\!\left(\frac{k}{2}\right)}{4\Gamma\!\left(\frac{k+1}{2}\right)}.
\eeq
We have $a_1=0$, in agreement with~(\ref{iden}),
whereas $a_2=1/\sqrt\pi$, $a_3=3\sqrt\pi/4$, $a_4=4/\sqrt\pi$,
$a_5=15\sqrt\pi/8$, and so on, are all positive.

In the case of the second moments,~(\ref{labs}) and~(\ref{idenexs}) yield the
exact expressions
\beq
\mean{M_n^2}=2n-2\mean{R_n},\qquad
\mean{R_n^2}=2n-3\mean{R_n},\
\eeq
with $\mean{R_n}$ being given by~(\ref{rcorr}).

An alternative presentation of some of the above results is given in~\ref{quick}.

\subsection{Simple Polya walk}
\label{simple}

To close, we consider the case of the simple Polya walk on the one-dimensional lattice
with unit spacing, with step length distribution
\beq
\rho(\eta)=\frac{1}{2}(\delta(\eta-1)+\delta(\eta+1)).
\eeq
Its variance reads $\sigma^2=1$, thus $D=1/2$.
The distribution $\rho(\eta)$ is not continuous,
so that some of the results derived so far must be revisited.

A peculiarity of the Polya walk is that the maximum and the number of records
coincide at all times $n$ for any given realization of the walk~\cite{revue,wms},
\beq
M_n=R_n,
\label{mrid}
\eeq
even in the presence of resetting~\cite{m2s2},
since any record breaking event corresponds to an increase of $M_n$ by one unit.

Let us first determine the steady-state distribution of the walker's position,
that we still denote as $f(x)$,
even though $x$ is now an integer random variable.
The Fourier transform $\F\rho(k)=\cos k$ is an even and $2\pi$-periodic function of $k$.
Equation~(\ref{fstatres}) yields
\beq
\F f(k)=\frac{r}{1-(1-r)\cos k}.
\eeq
We thus obtain
\beq
f(x)=\sqrt\frac{r}{2-r}\,\lam_{\rm st}^{-\abs{x}},
\label{fbin}
\eeq
with
\beq
\lam_{\rm st}=\frac{1+\sqrt{r(2-r)}}{1-r}.
\label{lam0def}
\eeq
The distribution~(\ref{fbin}) falls off exponentially, in agreement with~(\ref{fstatexp}).
The corres\-ponding decay rate,
\beq
\Kst=\ln\lam_{\rm st},
\label{k0def}
\eeq
is an increasing function of $r$,
behaving as $\Kst\approx\sqrt{2r}$ for $r\to0$, in accordance with~(\ref{ksca}),
and diverging as $\Kst\approx\ln(2/(1-r))$ as $r\to1$.

The maximum $M_n$ of the walk after $n$ steps takes integer values $M\ge0$.
Keeping in line with the definitions~(\ref{Fdef}) and~(\ref{qdef}),
the generating series
\beq
\w Q(z,x,M)=\sum_{n\ge0}Q_{n}(x,M)z^n,
\eeq
where both $x$ and $M$ are integers,
is non-zero only for $x\leq M$, where it obeys
\beq
\fl
\w Q(z,x,M)=1+rz\w Q(z,0,M)+\frac{\rz}{2}(\w Q(z,x-1,M)+\w Q(z,x+1,M))
\label{dqeq}
\eeq
(see~(\ref{rzdef})).
This equation has an exact solution of the form $\w Q(z,x,M)=A+B\lam^x$, which reads
\beq
\w Q(z,x,M)=\frac{1-\lam^{x-M-1}}{1-z+rz\lam^{-M-1}}\qquad(x\leq M),
\label{dq}
\eeq
with
\beq
\lam=\frac{1+\sqrt{1-\rz^2}}{\rz}.
\label{lamdef}
\eeq
The static $\lam_{\rm st}$ of~(\ref{lam0def}) is recovered for $z=1$.
Even though the step length distribution is not continuous,
the expression~(\ref{dq}) still has the structure of~(\ref{qfull}),
with
\beq
\w q(z,y)=\sum_{n\ge0}q_n(y)z^n=\frac{1-\lam_0^{-y-1}}{1-z}
\label{dg}
\eeq
and
\beq
\lam_0=\frac{1+\sqrt{1-z^2}}{z}.
\eeq
We have in particular (see~(\ref{fqser}))
\beq
\w F(z,M)=\frac{1-\lam^{-M-1}}{1-z+rz\lam^{-M-1}}.
\label{df}
\eeq
This expression encodes the exact distribution
\beq
f_{n}(M)=\prob(M_n=M)=F_{n}(M)-F_{n}(M-1)\qquad(M\ge0)
\eeq
of the maximum $M_n$ of the simple walk and of its number $R_n$ of records,
for all values of $n$ and $r$.
For $M=0$, the above equation is completed by the convention that $F_n(-1)=0$.

In the scaling regime where $n$ is large and $r$ is small,
setting $z=\e^{-s}$ where $s$ is also small,
we have $\ln\lam\approx\sqrt{2(s+r)}$, thus~(\ref{df}) becomes~(\ref{key}).
As a consequence,
all the results concerning the statistics of the maximum $M_n$
in the diffusive scaling regime derived in section~\ref{sca}
apply to the present situation of the simple Polya walk.

In order to go beyond the diffusive scaling regime,
we focus our attention on the mean values
\beq
\mean{M_n}=\mean{R_n}=\sum_{M\ge0}M\,f_{n}(M)=\sum_{M\ge0}(1-F_{n}(M)).
\eeq
The corresponding generating series reads
\beqa
\sum_{n\ge0}\mean{M_n}z^n
&=&\sum_{M\ge0}\left(\frac{1}{1-z}-\w F(z,M)\right)
\nonumber\\
&=&\frac{1-\rz}{1-z}\sum_{M\ge0}\frac{1}{rz+(1-z)\lam^{M+1}}.
\label{dqser}
\eeqa
We shall now successively discuss this result for $r=0$ and $r>0$.

\subsubsection*{In the absence of resetting.}

There,~(\ref{dqser}) becomes a geometric series.
We obtain
\beq
\sum_{n\ge0}\mean{M_n}z^n
=\sum_{n\ge0}\mean{R_n}z^n
=\frac{1}{2}\left(\frac{(1+z)^{1/2}}{(1-z)^{3/2}}-\frac{1}{1-z}\right),
\eeq
in agreement with earlier studies~\cite{revue,ziff,mms},
possibly up to a change of convention,
as the origin is not counted as a record in the present work.
The resulting expression of~$\mean{M_n}$ or $\mean{R_n}$ given in those
references is however rather complicated,
involving the hypergeometric function $_2F_1$.
We give here a simple explicit---and seemingly novel---expression of this mean value,
which depends on the parity of $n$ according to
\beq
\mean{M_{2k}}=\left(2k+\frac{1}{2}\right)b_k-\frac{1}{2},\qquad
\mean{M_{2k+1}}=\left(2k+1\right)b_k-\frac{1}{2},
\label{meo}
\eeq
where $b_n$ is defined in~(\ref{bdef}).
The asymptotic expansion of the above expression reads
\beq\label{eq:reduit}
\mean{M_n}=\mean{R_n}
=\sqrt\frac{2n}{\pi}\left(1+\frac{1}{4n}-\frac{1+2(-1)^n}{32n^2}+\cdots\right)-\frac{1}{2}.
\eeq
The second-order correction term keeps a trace of the parity effect evidenced
in~(\ref{meo}).

As already noticed in~\cite{ziff},
the leading behavior of $\mean{R_n}$ is $\sqrt2$ times smaller than the
universal formula~(\ref{mrmoms}) which holds for continuous symmetric distributions.
The occurrence of such a multiplicative factor is actually quite general
among discrete distributions, and more generally distributions having a discrete component.
For an arbitrary symmetric step length distribution,
the Spitzer formula~(\ref{spitzer}) for the mean increment
generalizes to~\cite{spitzer3,spitzerbook} (see also~\cite[ch.~XVIII]{feller2})
\beq
\mean{h_1}=E\sqrt{D},
\label{spitzergene}
\eeq
where the enhancement factor $E$ reads
\beq
E=\exp\left(\sum_{n\ge1}\frac{\prob(x_n=0)}{2n}\right)\ge1.
\label{edef}
\eeq
If the step length distribution is continuous,
the probability of having exactly $x_n=0$ is zero,
and so the enhancement factor $E$ is unity,
so that~(\ref{spitzer}) and~(\ref{mrequiv}) are recovered.
If the density $\rho(\eta)$ of the step length distribution contains delta functions,
either at the origin $(\eta=0)$ or at one or more pairs of symmetric positions $(\eta=\pm a)$,
the probability $\prob(x_n=0)$ might be non-zero, at least for some $n$,
so that one has generically $E>1$.
An interesting example is provided by the arithmetic distributions of the form
\beq
\rho(\eta)=\sum_{j=-J}^J f_j \delta(\eta-j),
\eeq
with $f_j=f_{-j}$ up to some finite range $J$.
Such distributions give rise to walks on the lattice of integers.
For this class of distributions, considered recently in~\cite{mms},
the Fourier transform $\F\rho(k)$ is an even and $2\pi$-periodic function of $k$.
We have
\beq
\prob(x_n=0)=\frac{1}{\pi}\int_0^\pi\F\rho(k)^n\,\dd k,
\eeq
and so
\beq
E=\exp\left(-\frac{1}{2\pi}\int_0^\pi\ln(1-\F\rho(k))\,\dd k\right).
\eeq

For arbitrary non-continuous symmetric step length distributions with finite variance,
inserting~(\ref{spitzergene}) into~(\ref{mrequiv2}),
we obtain a generalization of the asymptotic equivalence~(\ref{mrequiv}) in the form
\beq
M_n\approx E\sqrt{D}\,R_n.
\label{mrequivgene}
\eeq
As a consequence,
all the universal results for $M_n$ derived in section~\ref{sca}
for continuous distributions are unchanged,
including in the presence of a weak resetting,
whereas those concerning $R_n$ have to be modified
by taking the enhancement factor $E$ into account.
In particular, in the absence of resetting, $M_n$ and $R_n$ are still asymptotically
distributed according to half-Gaussian laws, with
\beq\label{eq:MnRn}
\mean{M_n}\approx2\sqrt\frac{Dn}{\pi},\qquad
\mean{R_n}\approx\frac{2}{E}\sqrt\frac{n}{\pi}.
\eeq
More importantly, throughout the diffusive scaling regime,
the analysis made in sections~\ref{scamoms} and~\ref{scafull} holds unchanged, with now
\beq
X=\frac{M}{\sqrt{Dn}}\quad\hbox{or}\quad X=\frac{ER}{\sqrt{n}}.
\eeq

For the simple Polya walk with weights $f_{\pm1}=1/2$, we have $D=1/2$ and $E=\sqrt{2}$,
so that~(\ref{eq:MnRn}) gives back~(\ref{eq:reduit}).
We have furthermore $\mean{h_1}=1$, as should be, since all increments are equal to unity,
so that the asymptotic equivalence $M_n\approx R_n$
is actually an identity (see~(\ref{mrid})).
In this sense the Polya walk is not generic.

In contrast, consider the symmetric walk of range 2 with weights $f_{\pm1}=f_{\pm2}=1/4$, hence
$D=5/2$, $E=\sqrt{5}-1$, and $\mean{h_1}=(5-\sqrt{5})/2$.
This example is now generic,
in the sense that the equivalence~(\ref{mrequivgene})
only holds asymptotically for late times.

\subsubsection*{In the presence of resetting.}

The asymptotic behavior of the mean maximum $\mean{M_n}$ of the simple Polya walk
for a fixed resetting probability can be studied as follows.
Setting $z=\e^{-s}$ where $s$ is small, while $r$ is kept fixed,
$\lam$ becomes $\lam_{\rm st}=\e^{\Kst}$ (see~(\ref{k0def})),
thereby~(\ref{dqser}) simplifies to
\beq
\sum_{n\ge0}\mean{M_n}z^n\approx\frac{r}{s}\sum_{M\ge0}\frac{1}{r+s\,\e^{\Kst
(M+1)}}.
\label{dser}
\eeq

A first estimate of the above series can be obtained by discarding its discrete nature
and replacing it by an integral over $M$.
We thus obtain
\beq
\sum_{n\ge0}\mean{M_n}z^n\approx\frac{\ln(r/s)}{\Kst s},
\eeq
hence
\beq
\mean{M_n}=\mean{R_n}\approx\frac{\ln nr}{\Kst}.
\label{dmrave}
\eeq
This logarithmic growth agrees with~(\ref{mave}).
Let us recall that~$\Kst$ depends on $r$ through~(\ref{lam0def}),~(\ref{k0def}).
The peculiarity of the Polya walk is that~(\ref{dmrave})
applies both to $\mean{M_n}$ and to $\mean{R_n}$.
The logarithmic growth law of $\mean{R_n}$
therefore involves the non-universal factor $\Kst$,
rather than the universal factor $\sqrt{r}$ entering~(\ref{rave})
for continuous step length distributions.

Furthermore, the logarithmic growth law~(\ref{dmrave}) is modulated by periodic
oscillations.
The more complete analysis of~(\ref{dser}) given in~\ref{series}
indeed yields the full asymptotic behavior (see~(\ref{sum1}),~(\ref{sum2}))
\beq
\mean{M_n}=\mean{R_n}\approx\frac{\ln nr+\gamma}{\Kst}-\frac{1}{2}+P(\ln nr),
\label{dmr}
\eeq
where $\gamma$ is Euler's constant,
whereas
\beq
P(v)=-\frac{2}{\Kst}\Re\sum_{m\ge1}\Gamma\!\left(-\frac{2\pi\ii m}{\Kst}\right)
\e^{2\pi\ii mv/\Kst}
\eeq
is an oscillating periodic function with zero average and period $\Kst$.

Periodic or log-periodic oscillations are usually met
in systems having a discrete symmetry, such as a discrete scale invariance
(see~\cite{sornette} for a review).
Here, they are a manifestation of the fact that $M_n=R_n$ is an integer random
variable.
The amplitude of the periodic function $P$ can be operationally defined as that
of its first harmonic ($m=1$), namely
\beq
A_1=\frac{2}{\Kst}\bigabs{\,\Gamma\!\left(\pm\frac{2\pi\ii}{\Kst}\right)}
=\sqrt{\frac{2}{\Kst\sinh(2\pi^2/\Kst)}}\sim\e^{-\pi^2/\Kst}.
\eeq
The periodic oscillations are therefore tiny,
except in the regime of a large resetting probability ($r\to1$),
where $\Kst$ diverges logarithmically,
so that the slope of the growth law~(\ref{dmrave}) becomes small.
For $r=0.5$, we have $\Kst\approx1.31695$ and $A_1\approx9.6947\ 10^{-4}$.
For $r=0.9$, we have $\Kst\approx2.99322$ and $A_1\approx4.2751\ 10^{-2}$.

\begin{figure}[!ht]
\begin{center}
\includegraphics[angle=0,width=.7\linewidth,clip=true]{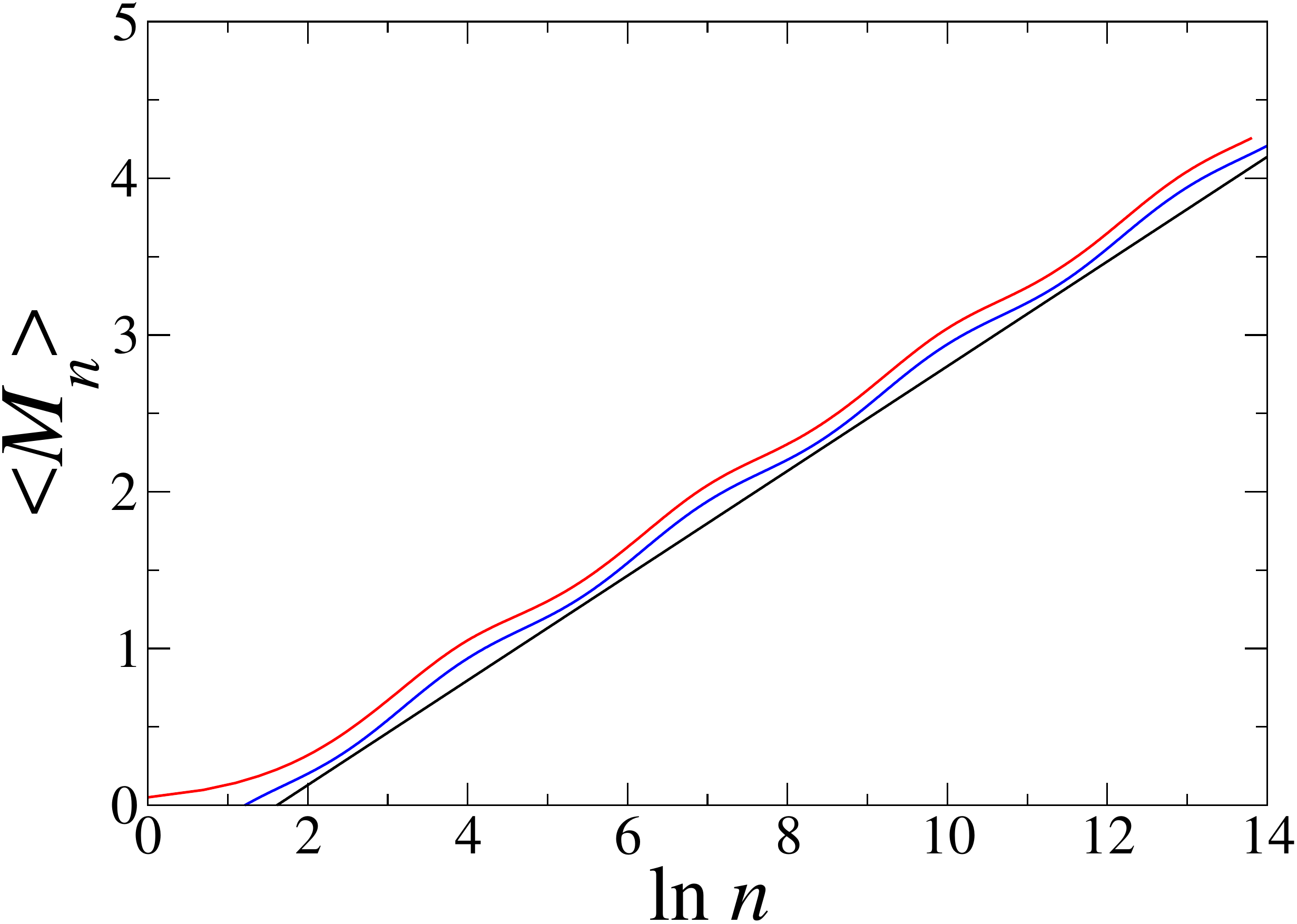}
\caption{\small
Red curve: mean value $\mean{M_n}$ against $\ln n$,
for Polya walks with a resetting probability $r=0.9$.
Blue curve (translated): full asymptotic expression~(\ref{dmr}).
The black line has the theoretical slope $1/\Kst\approx0.33408$.}
\label{arbin}
\end{center}
\end{figure}

The asymptotic expression~(\ref{dmr}) is compared with numerical data in figure~\ref{arbin},
showing $\mean{M_n}$ against $\ln n$,
as measured for simple Polya walks up to $n=10^6$, with a resetting probability
$r=0.9$.
The data (red) are observed to converge rapidly to the prediction~(\ref{dmr})
(blue),
including its oscillations.
The blue curve is slightly translated vertically for a better readability.
The black line has the theoretical slope $1/\Kst$.

\section{Discussion}
\label{disc}

In this work we have revisited the statistics of extremes and records
of symmetric random walks with stochastic resetting, thus complementing and extending
earlier studies on these topics.
The most salient of our findings are summarized below.

We have highlighted a diffusive scaling regime
for walks defined by symmetric step length distributions
with finite variance
and a weak resetting probability.
For continuous step length distributions, the maximum $M_n$ and the number $R_n$ of records
obey the remarkable equivalence~(\ref{mrequiv}),
implying that
these quantities become asymptotically proportional to each other,
even for single typical trajectories.
For step length distributions having a discrete component,
a result due to Spitzer allows to generalize the above equivalence to~(\ref{mrequivgene}),
where the dependence on the distribution is encoded in two parameters,
the diffusion coefficient $D$ and the enhancement factor $E$.
Within this framework,
the distributions of $M_n$ and $R_n$
obey scaling laws involving, besides $D$ and $E$,
a universal two-parameter scaling function $\Phi(X,u)$, as seen in section~\ref{scafull}.
As the mean number of resettings $u=nr$ is varied, the reduced distribution $\Phi(X,u)$
interpolates between a half-Gaussian law for $u\ll1$
and a Gumbel law for $u\gg1$.

We have also obtained various specific results
illustrating both quantitative and qualitative differences between
the statistics of $M_n$ and of $R_n$ beyond the diffusive scaling regime.
Exact results on the distribution of $M_n$ at finite times, obtained
for two particular step length distributions, corresponding respectively to
the symmetric exponential walk and the Polya lattice walk,
as well as a heuristic analysis of other distributions,
illustrate several facets of the statistics of extremes and records for random walks.
To take one noticeable example,
for symmetric walks with a fixed non-zero resetting probability $r$,
$\mean{R_n}$ grows logarithmically, according to the universal law
derived in~\cite{m2s2},
whereas $\mean{M_n}$ exhibits a variety of asymptotic behaviors.
For exponential and superexponential step length distributions,
$\mean{M_n}$ also grows logarithmically, according to~(\ref{mave}),
involving a non-universal amplitude $1/\Kst$ depending on $r$
and on the step length distribution.
For distributions falling off as a power law, as $\rho(\eta)\sim\abs{\eta}^{-(1+\theta)}$,
the typical value of $M_n$ grows as a power of time, as $M_n\sim n^{\alpha_r}$,
with $\alpha_r=1/\theta$, irrespective of the exponent $\theta>0$.
For usual random walks in the absence of resetting,
we have $\alpha_0=\alpha_r=1/\theta$ for $\theta<2$,
whereas $\alpha_0=1/2>\alpha_r$ in the diffusive case ($\theta>2$).

Yet many other features of the statistics of extremes and records
for random walks with stochastic resetting raise interesting open questions.
Investigating the survival probability of random walks and L\'evy flights with stochastic resetting
is a natural sequel to the present work~\cite{survie}.

\appendix

\section{Steady-state distribution of the walker's position}
\label{steady}

This Appendix is devoted to the distribution $f(x)$ of the walker's position
in the nonequilibrium steady state reached by the random walk with resetting
defined in~(\ref{def}).
Part of this material can be found in~\cite{emsrev}.
It is included here for completeness.
Notations are consistent with the body of the article, as far as possible.

The steady-state distribution $f(x)$ obeys the integral equation
\beq
f(x)=r\delta(x)+(1-r)\int_{-\infty}^\infty f(x-\eta)\rho(\eta)\dd\eta.
\label{fstateq}
\eeq
This equation has been obtained by conditioning on the last step of the walk,
which may consist of a resetting event.
It can be solved in Fourier space for any step length distribution.
With the notation~(\ref{foudef}) of Fourier transforms, we have
\beq
\F f(k)=\frac{r}{1-(1-r)\F\rho(k)}.
\label{fstatres}
\eeq
This expression can be expanded as
\beq
\F f(k)=\sum_{m\ge0}p_m\,\F\rho(k)^m,
\label{fstatexpand}
\eeq
where
\beq
p_m=r(1-r)^m\qquad(m\ge0)
\eeq
is the distribution of the age $m$ of the process,
i.e., the difference between the time at which the position $x$ is monitored
and the last resetting event.
More explicitly,~(\ref{fstatexpand}) yields
\beq
f(x)=r\delta(x)+f_\ct(x),
\label{fstatmix}
\eeq
where the delta function at the origin corresponds to $m=0$,
whereas the continuous component
\beq
f_\ct(x)=\sum_{m\ge1}p_m\,\stackunder{\underbrace{\rho(x)*\cdots*\rho(x)}}{m\;\mathrm{times}}
\eeq
receives contributions from all positive ages.

For diffusive walks, we have $\F\rho(k)\approx1-Dk^2$.
Expanding~(\ref{fstatres}), we obtain the expression of the variance of the
position
\beq
\int_{-\infty}^\infty x^2 f(x)\,\dd x=\frac{2(1-r)D}{r}.
\label{x2stat}
\eeq

The tails of the stationary distribution $f(x)$ are given by the following dichotomy.

For exponential and superexponential step length distributions,
i.e., distributions whose tails
are bounded by a decaying exponential of the form $\e^{-a\abs{\eta}}$,
the Fourier transform $\F\rho(k)$ is analytic in the strip $\abs{\Im k}<a$.
The steady-state distribution $f(x)$ decays exponentially as
\beq
f(x)\sim\e^{-\Kst\abs{x}}\qquad(x\to\pm\infty),
\label{fstatexp}
\eeq
where the decay rate $\Kst=-\ii k_0$ is the nearest pole of the analytic
continuation of~(\ref{fstatres}), obeying
\beq
(1-r)\F\rho(\ii\Kst)=1.
\label{kdef}
\eeq
In the weak-resetting regime, where $r$ is small,
the decay rate $\Kst$ itself becomes small and assumes the universal form
\beq
\Kst\approx\sqrt\frac{r}{D}.
\label{ksca}
\eeq
In this regime,~(\ref{fstatres}) boils down to
\beq
\F f(k)\approx\frac{r}{r+Dk^2},
\eeq
so that the bulk of the distribution $f(x)$ becomes the symmetric exponential
\beq
f(x)\approx\frac{\Kst}{2}\,\e^{-\Kst\abs{x}}.
\label{fstatsca}
\eeq
The variance of the position therefore scales as
\beq
\int_{-\infty}^\infty x^2 f(x)\,\dd x\approx\frac{2}{\Kst^2}\approx\frac{2D}{r},
\eeq
in quantitative agreement with~(\ref{x2stat}).

For subexponential step length distributions,
whose tails fall off more slowly than any exponential,
there is no open strip where the Fourier transform $\F\rho(k)$ is analytic.
In usual circumstances, $\F\rho(k)$ has an isolated singularity at the origin.
Denoting $\F\rho_\sg(k)$ its singular part as $k\to0$,~(\ref{fstatres}) yields
\beq
\F f_\sg(k)\approx\frac{1-r}{r}\,\F\rho_\sg(k),
\eeq
and so the stationary distribution $f(x)$
usually inherits the subexponential tails of the step length distribution:
\beq
f(x)\approx\frac{1-r}{r}\,\rho(x)\qquad(x\to\pm\infty).
\label{fsub}
\eeq

\section{Some identities involving Stirling numbers}
\label{stirling}

Stirling numbers~\cite{stir} play a central role in the combinatorics
of set partitions and of permutations
(see~\cite{knuth,GKP,FS} for comprehensive expositions).

Keeping notations consistent with the body of the paper,
the Stirling numbers of the first kind $\sone{k}{j}$ are defined by
\beq
\frac{\Gamma(z+k)}{\Gamma(z)}=z(z+1)\cdots(z+k-1)=\sum_{j=1}^k\sone{k}{j}z^j,
\label{soneid}
\eeq
and the Stirling numbers of the second kind $\stwo{k}{j}$ are defined by
\beq
z^k=\sum_{j=1}^k(-1)^{k-j}\stwo{k}{j}z(z+1)\cdots(z+j-1),
\eeq
so that we have the inversion formula
\beq
A_k=\sum_{j=1}^k\sone{k}{j}B_j
\Longleftrightarrow
B_k=\sum_{j=1}^k(-1)^{k-j}\stwo{k}{j}A_j.
\label{sinv}
\eeq

Let us begin by considering the series
\beq
S_k(y)=\sum_{n\ge0}n^ky^n.
\label{sabs}
\eeq
We have $S_0(y)=1/(1-y)$, as well as the differential recursion
\beq
S_k(y)=yS_{k-1}'(y),
\eeq
where the accent denotes a derivative.
Therefore,
\beq
S_k(y)=\frac{P_k(y)}{(1-y)^{k+1}},
\label{pabs}
\eeq
where $P_k(y)$ is a polynomial of degree $k$,
obeying the differential recursion
\beq
P_k(y)=y\left[kP_{k-1}(y)+(1-y)P_{k-1}'(y)\right].
\eeq
We have $P_0(y)=1$, $P_1(y)=y$, $P_2(y)=y(y+1)$, and so on.

Let us now form the combinations
\beqa
T_k(y)
&=&\sum_{j=1}^k\sone{k}{j}S_j(y)
\label{tdef}
\\
&=&\sum_{n\ge0}\sum_{j=1}^k\sone{k}{j}n^jy^n
\nonumber\\
&=&\sum_{n\ge0}n(n+1)\cdots(n+k-1)y^n.
\eeqa
The last expression was obtained by using~(\ref{soneid}).
The factor $(n+k-1)$ is present in~$T_k(y)$, but absent in $T_{k-1}(y)$.
This implies the differential recursion
\beq
T_k(y)=(k-1)T_{k-1}(y)+yT_{k-1}'(y),
\eeq
with $T_1(y)=S_1(y)=y/(1-y)^2$.
The solution of this recursion takes the simple form
\beq
T_k(y)=\frac{k!y}{(1-y)^{k+1}}.
\label{tabs}
\eeq
Using the inversion formula~(\ref{sinv}), we obtain
\beq
S_k(y)=\sum_{j=1}^k(-1)^{k-j}\stwo{k}{j}T_j(y)
\eeq
and
\beq
P_k(y)=y\sum_{j=1}^k\stwo{k}{j}j!(y-1)^{k-j}\qquad(k\ge1).
\eeq

Some of the above identities can be applied to the setting of
section~\ref{se}.
Comparing~(\ref{qabs}) to~(\ref{sabs}) and~(\ref{labs}) to~(\ref{tabs}), we
obtain
\beq
\sr{k}(z)=\frac{S_k(1-\nu_0)}{\nu_0},\qquad
\sm{k}(z)=\frac{T_k(1-\nu_0)}{\nu_0}.
\eeq
Equation~(\ref{tdef}) then yields
\beq
\sm{k}(z)=\sum_{j=1}^k\sone{k}{j}\sr{j}(z),
\eeq
and finally
\beq
\mean{M_n^k}=\sum_{j=1}^k\sone{k}{j}\mean{R_n^j}.
\eeq
This is the first identity in~(\ref{idens}).
The second one is a consequence of the inversion formula~(\ref{sinv}).

\section{Alternative derivation of some of the results of~section~\ref{se}}
\label{quick}

The starting point is the observation that for a Laplace step length
distribution (\ref{sedef}) there is decoupling between the distribution of the
increments $f_{h_1}(h)$ and that of the record number $p_n(R)$~\cite{revue},
\be
\frac{\dd}{\dd h}\prob(h_1<h,R_n=R)=\e^{-h}p_n(R),
\ee
where $p_n(R)$ is given by~(\ref{pnR}).
This property allows to recover in a straightforward way the expression of the
density of the maximum $M_n$, already known thanks to (\ref{gexp}) to be given by
\beq\label{eq:fMn}
\sum_{n\ge0} f_n(M)\, z^n =\tilde
q(z)\delta(M)+\frac{1-\sqrt{1-z}}{\sqrt{1-z}}\e^{-M\sqrt{1-z}},
\eeq
where the first term in the r.h.s.~corresponds to $R_n=0$.
Since $M_n$ is given by the sum (see~(\ref{eq:Mn}))
\beq
M_n=h_1+h_2+\cdots+h_{R_n},
\eeq
we have
\beq\label{eq:start}
f_n(M)=\,\stackunder{\underbrace{q_n\,\delta(M)}}{R=0}+\sum_{R=1}^{n}p_n(R)(f_{h_1}\ast)^R(M),
\eeq
with, for the convolution of $R$ times the density $f_{h_1}(h)=\e^{-h}$,
\beq
(f_{h_1}\ast)^R(M)=\e^{-M}\frac{M^{R-1}}{(R-1)!},\qquad R\ge1.
\eeq
The generating function of the two sides of (\ref{eq:start}) yields (\ref{eq:fMn})
back.

We are now in position to compute the moments of $M_n$.
Indeed, for $k\ge1$,
\beqa
\fl
\mean{M_n^k}=
\int_0^\infty f_n(M)M^k\,\dd M
&=&\sum_{R=1}^{n}p_n(R)
\int_0^\infty(f_{h_1}\ast)^R(M)\,M^k\,\dd M
\nonumber\\
\label{eq:momentM}
&=&\sum_{R=1}^{n}p_n(R)\frac{\Gamma(k+R)}{\Gamma(R)}.
\eeqa
Using~(\ref{soneid})
we conclude that
\beq
\mean{M_n^k}
=\sum_{R=1}^{n}p_n(R)\sum_{j=1}^k\sone{k}{j}R^j
=\sum_{j=1}^k\sone{k}{j}\mean{R_n^j},
\eeq
which is the first equality in~(\ref{idens}).

Likewise,
\beq
\fl
\sum_{n\ge0}z^n\mean{M_n^k}
=\frac{1-\sqrt{1-z}}{\sqrt{1-z}}\int_0^\infty\e^{-M\sqrt{1-z}}\,M^k\,\dd M
=\frac{(1-\sqrt{1-z})\,k!}{(\sqrt{1-z})^{k+2}},
\eeq
which can also be obtained using (\ref{eq:momentM}) and (\ref{psae}),
\beq
\fl
\sum_{n\ge0}z^n\mean{M_n^k}=\sum_{R=1}^{n}\frac{(1-\nu_0)^R}{\nu_0}
\frac{\Gamma(k+R)}{\Gamma(R)}
=\frac{(1-\sqrt{1-z})\,k!}{(\sqrt{1-z})^{k+2}}.
\eeq
Equation (\ref{labs}) is thus recovered.

\section{Asymptotic behavior of the series involved in~(\ref{dser})}
\label{series}

This appendix is devoted to the asymptotic behavior as $\eps\to0$ of the series
\beq
S(\eps)=\sum_{M\ge0}\frac{1}{1+\eps\,\e^{\Kst(M+1)}}.
\eeq
This series is involved in~(\ref{dser}), which reads
\beq
\sum_{n\ge0}\mean{M_n}z^n\approx\frac{S(s/r)}{s}.
\label{conser}
\eeq

Following the line of thought of~\cite{dil},
let us introduce the Mellin transform
\beqa
T(p)
&=&\int_0^\infty\eps^{p-1}\,S(\eps)\dd\eps
\nonumber\\
&=&\sum_{M\ge0}\int_0^\infty\frac{\eps^{p-1}}{1+\eps\,\e^{\Kst(M+1)}}\,\dd\eps
\nonumber\\
&=&\sum_{M\ge0}\e^{-p\Kst(M+1)}\int_0^\infty\frac{u^{p-1}}{1+u}\,\dd u
\nonumber\\
&=&\frac{1}{\e^{p\Kst}-1}\,\frac{\pi}{\sin p\pi}\qquad(0<\Re p<1).
\eeqa
The third line is obtained by setting $\eps=\e^{-\Kst(M+1)}u$,
and the fourth one by performing separately the geometric sum over $M$ and the
integral over $u$.

The inverse Mellin formula reads
\beq
S(\eps)=\int\frac{\dd p}{2\pi\ii}\,\eps^{-p}\,T(p).
\eeq
The leading behavior of $S(\eps)$ as $\eps\to0$ is dictated by the rightmost
poles of $T(p)$
to the left of the integration contour.
There are an infinity of poles such that $\Re p=0$.
The double pole at $p=0$ yields
\beq
S_1(\eps)=-\frac{\ln\eps}{\Kst}-\frac{1}{2},
\eeq
whereas the simple poles at $p=2\pi\ii m/\Kst$ for $m=\pm1,\pm2,\dots$ yield
\beq
S_2(\eps)=-\frac{\ii\pi}{\Kst}\sum_{m\ne0}\frac{\e^{-2\pi\ii
m\ln\eps/\Kst}}{\sinh(2\pi^2m/\Kst)}.
\eeq

We have therefore
\beq
S(\eps)\approx S_1(\eps)+S_2(\eps),
\eeq
up to terms of order $\eps$.
Inserting this into~(\ref{conser}) and performing the inverse Laplace transform
from $s$ to $n$,
we obtain
\beq
\mean{M_n}\approx\frac{\ln nr+\gamma}{\Kst}-\frac{1}{2}+P(\ln nr),
\label{sum1}
\eeq
where $\gamma$ is Euler's constant,
whereas
\beq
P(v)=-\frac{2}{\Kst}\Re\sum_{m\ge1}\Gamma\!\left(-\frac{2\pi\ii
m}{\Kst}\right)\e^{2\pi\ii mv/\Kst}
\label{sum2}
\eeq
is an oscillating periodic function with zero average and period $\Kst$.

\section*{References}

\bibliography{paper.bib}

\providecommand{\newblock}{}
\begin{thebibliography}{10}
\expandafter\ifx\csname url\endcsname\relax
  \def\url#1{{\tt #1}}\fi
\expandafter\ifx\csname urlprefix\endcsname\relax\def\urlprefix{URL }\fi
\providecommand{\eprint}[2][]{\url{#2}}

\bibitem{emsrev}
Evans M~R, Majumdar S~N and Schehr G 2020 {\em J. Phys. A: Math. Theor.\/} {\bf
  53} 193001

\bibitem{m2s2}
Majumdar S~N, Mounaix P, Sabhapandit S and Schehr G 2022 {\em J. Phys. A: Math.
  Theor.\/} {\bf 55} 034002

\bibitem{em}
Evans M~R and Majumdar S~N 2011 {\em Phys. Rev. Lett.\/} {\bf 106} 160601

\bibitem{mmss}
Majumdar S~N, Mori F, Schawe H and Schehr G 2021 {\em Phys. Rev. E\/} {\bf 103}
  022135

\bibitem{revue}
Godr\`eche C, Majumdar S~N and Schehr G 2017 {\em J. Phys. A: Math. Theor.\/}
  {\bf 50} 333001

\bibitem{hopf}
Hopf E 1934 {\em Mathematical Problems of Radiative Equilibrium\/} (Cambridge:
  Cambridge University Press)

\bibitem{chandra}
Chandrasekhar S 1960 {\em Radiative Transfer\/} (New York: Dover)

\bibitem{LA}
Lawrie J~B and Abrahams I~D 2007 {\em J. Eng. Math.\/} {\bf 59} 351--358

\bibitem{spitzer1}
Spitzer F 1957 {\em Duke Math. J.\/} {\bf 24} 327--343

\bibitem{spitzer2}
Spitzer F 1960 {\em Duke Math. J.\/} {\bf 27} 363--372

\bibitem{ivanov}
Ivanov V~V 1994 {\em Astron. Astrophys.\/} {\bf 286} 328--337

\bibitem{sparre53}
\protect{Sparre Andersen} E 1953 {\em Math. Scand.\/} {\bf 1} 263--285

\bibitem{sparre54}
\protect{Sparre Andersen} E 1954 {\em Math. Scand.\/} {\bf 2} 194--222

\bibitem{feller2}
Feller W 1971 {\em An Introduction to Probability Theory and its
  Applications\/} 2nd ed vol~2 (New York: Wiley)

\bibitem{blackwell}
Blackwell D 1953 {\em Pacific J. Math.\/} {\bf 3} 315--320

\bibitem{spitzer3}
Spitzer F 1960 {\em Trans. Amer. Math. Soc.\/} {\bf 94} 150--169

\bibitem{spitzerbook}
Spitzer F 2001 {\em Principles of Random Walk\/} (New York: Springer)

\bibitem{gmsprl}
Godr\`eche C, Majumdar S~N and Schehr G 2016 {\em Phys. Rev. Lett.\/} {\bf 117}
  010601

\bibitem{ustc}
Godr\`eche C and Luck J~M Unpublished

\bibitem{milne}
Milne E~A 1921 {\em Monthly Notices Roy. Astron. Soc.\/} {\bf 81} 361--375

\bibitem{mounaix17}
Majumdar S~N, Mounaix P and Schehr G 2017 {\em J. Phys. A: Math. Theor.\/} {\bf
  50} 465002

\bibitem{wms}
Wergen G, Majumdar S~N and Schehr G 2012 {\em Phys. Rev. E\/} {\bf 86} 011119

\bibitem{bachelier}
Bachelier L 1901 {\em Ann. Sci. Ecole Normale Sup.\/} {\bf 18} 143--209

\bibitem{levy}
L\'evy P 1940 {\em Compositio Math.\/} {\bf 7} 283--339

\bibitem{borodin}
Borodin A~N and Salminen P 1996 {\em Handbook of Brownian Motion - Facts and
  Formulae\/} (Basel: Birkh\"auser)

\bibitem{ziff}
Majumdar S~N and Ziff R~M 2008 {\em Phys. Rev. Lett.\/} {\bf 101} 050601

\bibitem{singh}
Singh P and Pal A 2021 {\em Phys. Rev. E\/} {\bf 103} 052119

\bibitem{glrenew}
Godr\`eche C and Luck J~M 2001 {\em J. Stat. Phys.\/} {\bf 104} 489--524

\bibitem{cm}
Comtet A and Majumdar S~N 2005 {\em J. Stat. Mech.\/}  P06013

\bibitem{mcz}
Majumdar S~N, Comtet A and Ziff R~M 2006 {\em J. Stat. Phys.\/} {\bf 122}
  833--856

\bibitem{mms}
Mounaix P, Majumdar S~N and Schehr G 2020 {\em J. Phys. A: Math. Theor.\/} {\bf
  53} 415003

\bibitem{sornette}
Sornette D 1998 {\em Phys. Rep.\/} {\bf 297} 239--270

\bibitem{survie}
Godr\`eche C and Luck J~M 2022 Survival probability of random walks and
  \protect{L\'evy} flights with stochastic resetting (\textit{Preprint}
  \eprint{arXiv:2204.07392})

\bibitem{stir}
Stirling J 1730 {\em Methodus Differentialis\/} (London: Bowyer)

\bibitem{knuth}
Knuth D~E 1968 {\em The Art of Computer Programming\/} (New York:
  Addison-Wesley)

\bibitem{GKP}
Graham R~L, Knuth D~E and Patashnik O 1989 {\em Concrete Mathematics: A
  Foundation for Computer Science\/} (Reading, MA: Addison-Wesley)

\bibitem{FS}
Flajolet P and Sedgewick R 2009 {\em Analytic Combinatorics\/} (Cambridge:
  Cambridge University Press)

\bibitem{dil}
Derrida B, Itzykson C and Luck J 1984 {\em Commun. Math. Phys.\/} {\bf 94}
  115--132

\end{thebibliography}

\end{document}